\documentclass[AMA,STIX1COL]{WileyNJD-v2}

\articletype{Article Type}%

\received{26 April 2016}
\revised{6 June 2016}
\accepted{6 June 2016}

\begin{document}
\title{\ Long-Term Dagum-PVF Frailty Regression Model: Application in Health Studies\protect}
%\title{\ Long-term Dagum-PVF frailty regression model: Applications to datasets of a maternal population with COVID-19 and of malignant skin neoplasms \protect} %\thanks{Long-Term Dagum PVF Frailty Regression Model: An application to severe COVID-19 in pregnant and postpartum women dataset.}
 
\author[1,2]{Agatha S Rodrigues*}

\author[1]{Patrick Borges}

\authormark{Rodrigues \& Borges}

\address[1]{\orgdiv{Department of Statistics}, \orgname{Federal University of Espirito Santo}, \orgaddress{\state{Espirito Santo}, \country{Brazil}}}

\address[2]{\orgdiv{Division of Clinical Obstetrics, Hospital das Clinicas HCFMUSP, Faculty of Medicine}, \orgname{University of São Paulo}, \orgaddress{\state{Sao Paulo}, \country{Brazil}}}

\corres{*Corresponding author\\ \email{agatha.rodrigues@ufes.br}}

\presentaddress{Federal University of Espirito Santo \\
Av. Fernando Ferrari, 514 \\
Goiabeiras \\
Vitoria - ES \\
29075-910 \\
Brazil}

\abstract[Summary]{Survival models incorporating cure fractions, commonly known as cure fraction models or long-term survival models, are widely employed in epidemiological studies to account for both immune and susceptible patients in relation to the failure event of interest under investigation. In such studies, there is also a need to estimate the unobservable heterogeneity caused by prognostic factors that cannot be observed. Moreover, the hazard function may exhibit a non-monotonic form, specifically, an unimodal hazard function. In this article, we propose a long-term survival model based on the defective version of the Dagum distribution, with a power variance function (PVF) frailty term introduced in the hazard function to control for unobservable heterogeneity in patient populations, which is useful for accommodating survival data in the presence of a cure fraction and with a non-monotone hazard function. The distribution is conveniently reparameterized in terms of the cure fraction, and then associated with the covariates via a logit link function, enabling direct interpretation of the covariate effects on the cure fraction, which is not usual in the defective approach. It is also proven a result that generates defective models induced by PVF frailty distribution. We discuss maximum likelihood estimation for model parameters and evaluate its performance through Monte Carlo simulation studies. Finally, the practicality and benefits of our model are demonstrated through two health-related datasets, focusing on severe cases of COVID-19 in pregnant and postpartum women and on patients with malignant skin neoplasms.}


\jnlcitation{\cname{%
\author{Rodrigues A.},
\author{Borges P.}} (\cyear{2023}),
\ctitle{Long-Term Dagum-PVF Frailty Regression Model: Application in Health Studies}.}

\maketitle

%\footnotetext{\textbf{Abbreviations:} ANA, anti-nuclear antibodies; APC, antigen-presenting cells; IRF, interferon regulatory factor}

\section{Introduction}\label{introduction}

In survival studies, two major regression models stand out: the Cox proportional hazards model \cite{cox1972regression} and the accelerated failure rate model \cite{prentice1978linear}. These models work under the assumption that every individual will inevitably experience the event of interest within the observation period. However, real-life scenarios often involve individuals who don't experience the event, either due to being cured or developing immunity. The proportion of such individuals who are cured has received significant attention in traditional survival analysis, particularly in clinical studies \citep{lawless2011statistical}.

To address this situation, cure fraction models, also known as long-term survival models, are commonly used. They account for both individuals susceptible to the event and those who are cured or immune. The standard mixture model, initially proposed by Boag \cite{boag1949maximum} and extended by Berkson and Gage \cite{berkson1952survival}, is widely employed. This model defines the survival function as \(S(t) = p_{0} + \big(1-p_{0}\big)S_{1}(t)\), where \(S_1(\cdot)\) represents the survival function of individuals susceptible to the event, and \(p_{0}\) is the cure fraction. The choice of \(S_1(\cdot)\) can be based on distributions like Weibull, log-normal, log-logistic, or others. This approach enables the exploration of covariate effects on both the cure fraction and the survival time of individuals susceptible to the event.

Another approach involves models based on defective distributions \cite{balka2009review, balka2011bayesian}. These distributions alter one parameter's parametric space, ensuring that the survival function remains bounded between 0 and 1 as time approaches infinity. A notable advantage of defective distributions is their ability to model a cure fraction without introducing additional parameters, unlike the standard mixture model. Instead, the cure fraction relies on the parameters of the chosen distribution. Several studies have explored various distributions in their defective forms, such as Gompertz, inverse Gaussian, and extensions of the Marshall–Olkin distribution \cite{rocha2016two, rocha2017new, scudilio2019defective, calsavara2019defective, calsavara2019zero}.

A common challenge in survival data analysis is individual heterogeneity arising from unobservable factors like genetics, environment, or stochastic effects. To tackle this, frailty models introduce a random effect representing unobserved heterogeneity among individuals \cite{hougaard1994heterogeneity}. This random effect can be added or introduced multiplicatively in the baseline hazard function, extending the Cox proportional hazards model \cite{calsavara2017frailty}. Although frailty models have been explored in long-term survival models, our work introduces a new regression model: the Dagum-PVF model for analyzing survival data with long-term survivors.

The Dagum-PVF model combines the defective version of the Dagum distribution \cite{dagum1977} with a power variance function frailty term. This model offers several advantages: i) estimates the cure fraction without requiring an additional parameter; ii) allows for the direct interpretation of covariate effects on the cure fraction, which is not common in defective approach; iii) does not presuppose the existence of a predetermined cure rate, that is, the data indicates whether there is a cure or not depending on the value of the shape parameter; iv) allows the existence of cure for a subgroup and no cure for another; v) accommodates non-monotonic hazard functions. 

To showcase the versatility and practicality of the Dagum-PVF model, we apply it to two real first-analyzed datasets. One is related to the cases of pregnant and postpartum women (maternal population) hospitalized with severe acute respiratory syndrome (SARS) confirmed by COVID-19 in Brazil. The second one is a population-based dataset of incident cases of malignant neoplasms of skin in the state of São Paulo, Brazil.
 
The structure of this article is as follows: Section \ref{background} provides the background information. Section \ref{dagum_PVF} is devoted to the development of the defective models induced by PVF frailty, resulting in the long-term Dagum-PVF frailty model. The inferential process is outlined in Section \ref{inference}. Section \ref{simulation} contains the presentation of simulation studies. In Section \ref{application}, you will find the results of the two applications. We conclude with final considerations in Section \ref{final_considerations}.

\section{Background}
\label{background}

In this section, we present the Dagum distribution, frailty model, as well as the PVF frailty model. The Dagum distribution is
useful for modeling data with the possibility of long-term survivors in
the population.

\subsection{Dagum distribution}

The three-parameters type I Dagum distribution was introduced by Dagum \cite{dagum1977} and the survival function is given by
\begin{equation}
\label{eq-1}
S(t\mid \kappa, \alpha,\beta,\gamma)= 1-\left[\kappa + \left(\frac{t}{\beta}\right)^{-\alpha}\right]^{-\gamma}, \quad t > 0,
\end{equation}
where $\kappa=1$,  $\alpha > 0$ and $\gamma > 0$ are shape parameters and $\beta > 0$ is a scale parameter. 

Based on the type I Dagum distribution \eqref{eq-1},  Martinez and Achcar  \cite{martinez2018new} introduced a new defective distribution by considering $\kappa = \theta^{-1}$, with $0 < \theta \leq 1$, and $\gamma = 1$. Thus, the defective Dagum (DD) distribution  has density and survival functions, respectively, given by
\begin{equation*}
%\label{dens_surv}
f_{DD}(t\mid\alpha,\beta,\theta)= \alpha\beta\theta^2\frac{t^{-(\alpha+1)}}{(\beta + \theta t ^{-\alpha})^2},  
\end{equation*}
and
\begin{equation}
\label{dd_surv}
S_{DD}(t\mid\alpha,\beta,\theta)= \frac{\beta + \theta t^{-\alpha}-\theta\beta}{\beta + \theta t^{-\alpha}}, \quad t > 0.
\end{equation}
The cure fraction is:
\begin{equation*}
\lim_{t\rightarrow\infty}S_{DD}(t\mid\alpha,\beta,\theta)= 1 - \theta =p_{0}(\theta)\in(0,1),
\end{equation*}
 with $S_{DD}(t\mid\alpha,\beta,\theta)$ presented in \eqref{dd_surv}.
Thus, the cure fraction depends only on the parameter $0 < \theta < 1$, which provides a great advantage for the DD distribution in relation to other defective distributions found in the literature, that depend on all distribution parameters in general. The DD distribution becomes a proper distribution if $\theta = 1$, that is,  $\lim_{t\rightarrow\infty}S_{DD}(t\mid\alpha,\beta,\theta=1)=0$.

The hazard function of the DD distribution is given by
\begin{equation}
\label{dd_hazard}
h(t\mid\alpha,\beta,\theta)= \frac{\alpha\beta\theta^2t^{-(\alpha+1)}}{(\beta + \theta t ^{-\alpha})(\beta + \theta t^{-\alpha}-\theta\beta)}.
\end{equation}

\subsection{Frailty model}
In practical problems, it is very common that important features for the study are not observed, such as smoke exposure, genetics or epigenetics effects and other information that make the exposition varies from different individuals \cite{gazon2022nonproportional}. 
The non-observation of these variables can lead to an unobserved heterogeneity among individuals and ignoring unobserved heterogeneity may lead to inaccurate estimates \cite{calsavara2017frailty}.

To overcome that, a random effect that has a multiplicative effect on the baseline hazard function is considered \cite{wienke2010frailty}. This random variable is termed by Vaupel \textit{et al.} \cite{vaupel1979impact} as frailty. 
 Let $Z$ be an unobservable, nonnegative random variable and the conditional hazard function of the frailty model is defined by
 \begin{equation*}
h(t\mid z)= z h_0(t),
\end{equation*}
where $z$ represents the value of $Z$ and $h_0(\cdot)$ is the baseline hazard function, that is common to all individuals. The higher the value of $Z$ the greater the chance of an event occurring. 

The conditional survival function is given by
 \begin{equation}
 \label{eq-suv_frailty}
S(t\mid z)= \exp\{-zH_0(t)\},
\end{equation}
in which $H_0(t)=-\log[S_0(t)]$ is the baseline cumulative hazard function. 

The unconditional survival function is determined by integrating $S(t\mid z)$ in \eqref{eq-suv_frailty} in relation to the distribution of $Z$, in which $f_Z(z)$ is the respective density function. Thus, 
 \begin{equation}
 \label{eq-suv_frailty_unc}
S(t)= {\rm{E}}_{Z}[S(t\mid z)]=\int_{0}^{\infty} \exp\left[-zH_0(t)\right]f_z(z)dz  =  \hspace{0.05cm} L_z[H_0(t)],
\end{equation}
where $L_z[H_0(t)]$ is the Laplace transform of the frailty distribution. The unconditional hazard function is given by
 \begin{equation*}
% \label{eq-haz_frailty_unc}
h(t)= -\frac{d}{dt}log[S(t)]   = -\frac{h_0(t)L'_z[H_0(t)]}{L_z[H_0(t)]},
\end{equation*}
where $L'_z[H_0(t)]$ is the first derivative of the Laplace transform function with respect to the time $t$.

%From (6) and (7), natural candidates for the frailty distributions are those distributions possessing a Laplace transform function on the closed form because it facilitates the use of the traditional ML methods for parameter estimation (Wienke,2010). However, when the frailty distribution has no Laplace transform on the closed-form, numerical, or stochastic integration methods such as Gaussian quadrature and Monte Carlo simulation, among others, can be used to approximate the integral given in (6) (Balakrishnan & Peng, 2006; Hougaard, 2012; Robert & Casella, 2013). Pickles and Crouchley (1995) has mentioned that computational convenience is a factor that must be taken into account in practice when considering the frailty distribution in the modeling of univariate and multivariate survival data. Therefore, in this paper, we assume that the WL distribution is the frailty distribution (see the following sections).

\subsection{PVF Frailty model}

Let $Z$ be a random variable following a PVF distribution with parameters $\mu, \psi$, and $\gamma$ so that the density function can be written as \citep{wienke2010frailty}
\begin{eqnarray}
f_Z(z\mid \mu,\psi,\gamma)&=&\exp\left[-\psi(1-\gamma)\left(\frac{z}{\mu}-\frac{1}{\gamma}\right)\right]\frac{1}{\pi}\sum_{k=1}^{\infty}(-1)^{k+1}\frac{[\psi(1-\gamma)]^{k(1-\gamma)}\mu^{k\gamma}\Gamma(k\gamma+1)}{\gamma^k k!}z^{-k\gamma-1}\nonumber \\
&& ~~~~  \times  \sin(k\gamma \pi), \nonumber\label{densidadepvf}
\end{eqnarray}
where $\mu>0$, $\psi >0$ and $0<\gamma\leq 1$. 

In order to make sure that the model is identifiable, the restriction $\rm{E}(Z|\mu,\psi,\gamma)=\mu=1$  \citep{wienke2010frailty}. Consequently  the $\rm{Var}(Z|\mu,\psi,\gamma)=\mu^2/\psi=\sigma^2$, where $\sigma^2$ is interpreted as the measure of unobserved heterogeneity. %With this restriction, the results PVF parameters are $\gamma$ and $\sigma^2$.

The unconditional survival, obtained as presented in the Equation \eqref{eq-suv_frailty_unc}, and density functions in the PVF frailty model are expressed, respectively, by
\begin{eqnarray}
\label{surv_pvf}
\displaystyle S(t\mid \gamma,\sigma^2)=\exp\left\{\frac{1-\gamma}{\gamma \sigma^2}\left[1-\left(1-\frac{\sigma^2 \log[S_0(t)]}{1-\gamma}\right)^{\gamma}\right]\right\}
\end{eqnarray}
and
\begin{eqnarray}
 \label{dens_pvf}
	\displaystyle f(t\mid \gamma,\sigma^2)=h_0(t)\left(1-\frac{\sigma^2\log[S_0(t)]}{1-\gamma}\right)^{\gamma-1}\exp\left\{\frac{1-\gamma}{\gamma \sigma^2}\left[1-\left(1-\frac{\sigma^2 \log[S_0(t)]}{1-\gamma}\right)^{\gamma}\right]\right\}.
	\end{eqnarray}

As shown, the PVF distribution provides an algebraic treatment of the closed form for the unconditional survival function. Besides, it is a flexible model in the sense that it includes many other frailty models as special cases. The gamma frailty model is obtained if $\gamma \rightarrow 0$; and, in the case of $\gamma = 0.5$, the inverse Gaussian distribution is derived. The positive stable is a special case of the PVF distribution with some asymptotic considerations. 

\section{Long term Dagum-PVF frailty model}
\label{dagum_PVF}
Within section, we initially introduce a new class of defective models that involve the the PVF  frailty term. An essential outcome highlighted in this paper reveals that if the survival function $S_0(t)$ in Equation \eqref{surv_pvf} is defective, then the survival function $S(t\mid \gamma,\sigma^2)$ also originates from a defective distribution. This is formally expressed below:
\begin{theorem}
\label{ter1}
If $S_0(t)$ is a survival function of a defective distribution, then $S(t\mid \gamma,\sigma^2)$ in Equation \eqref{surv_pvf} is also a survival function from a defective distribution.
\end{theorem}
The complete proof for Theorem~\ref{ter1} is provided in the Appendix.

\vspace{0.2cm}

Now, we proceed to establish the long-term Dagum-PVF frailty model by replacing $H_0(t)$ and $h_0(t)$ in equations \eqref{surv_pvf} and \eqref{dens_pvf} by $H_{DD}(t\mid \alpha, \beta, \theta) = -\log[S_{DD}(t\mid \alpha, \beta, \theta)]$, where $S_{DD}(t\mid \alpha, \beta, \theta)$ is presented in Equation \eqref{dd_surv}, and $h_{DD}(t\mid \alpha, \beta, \theta)$ presented in Equation \eqref{dd_hazard}, respectively. Thus, the defective Dagum PVF (DD-PVF) frailty model survival and density functions are, respectively, given by

\begin{eqnarray}
\label{surv_dagum_pvf}
\displaystyle S(t\mid \mbox{\boldmath{$\vartheta$}})=\exp\left(\frac{1-\gamma}{\gamma \sigma^2}\left\{1-\left[1-\frac{\sigma^2}{1-\gamma}\log\left(\frac{\beta + \theta t^{-\alpha}-\theta\beta}{\beta + \theta t^{-\alpha}}\right)\right]^{\gamma}\right\}\right)
\end{eqnarray}
and
\begin{eqnarray}
 \label{dens_dagum_pvf}
	\displaystyle f(t\mid \mbox{\boldmath{$\vartheta$}})=\frac{\alpha\beta\theta^2t^{-(\alpha+1)}}{(\beta + \theta t ^{-\alpha})(\beta + \theta t^{-\alpha}-\theta\beta)}\left[1-\frac{\sigma^2}{1-\gamma}\log\left(\frac{\beta + \theta t^{-\alpha}-\theta\beta}{\beta + \theta t^{-\alpha}}\right)\right]^{\gamma-1}\nonumber \\
 ~~~~~~ ~~~ ~~~~ \times \exp\left(\frac{1-\gamma}{\gamma \sigma^2}\left\{1-\left[1-\frac{\sigma^2}{1-\gamma}\log\left(\frac{\beta + \theta t^{-\alpha}-\theta\beta}{\beta + \theta t^{-\alpha}}\right)\right]^{\gamma}\right\}\right),
 \end{eqnarray}
where $\mbox{\boldmath{$\vartheta$}} = (\alpha, \beta, \theta, \gamma,\sigma^2)^{\top} $. 

\begin{figure}[h!]
	\centering
	\begin{center}
		\includegraphics[scale = 0.75]{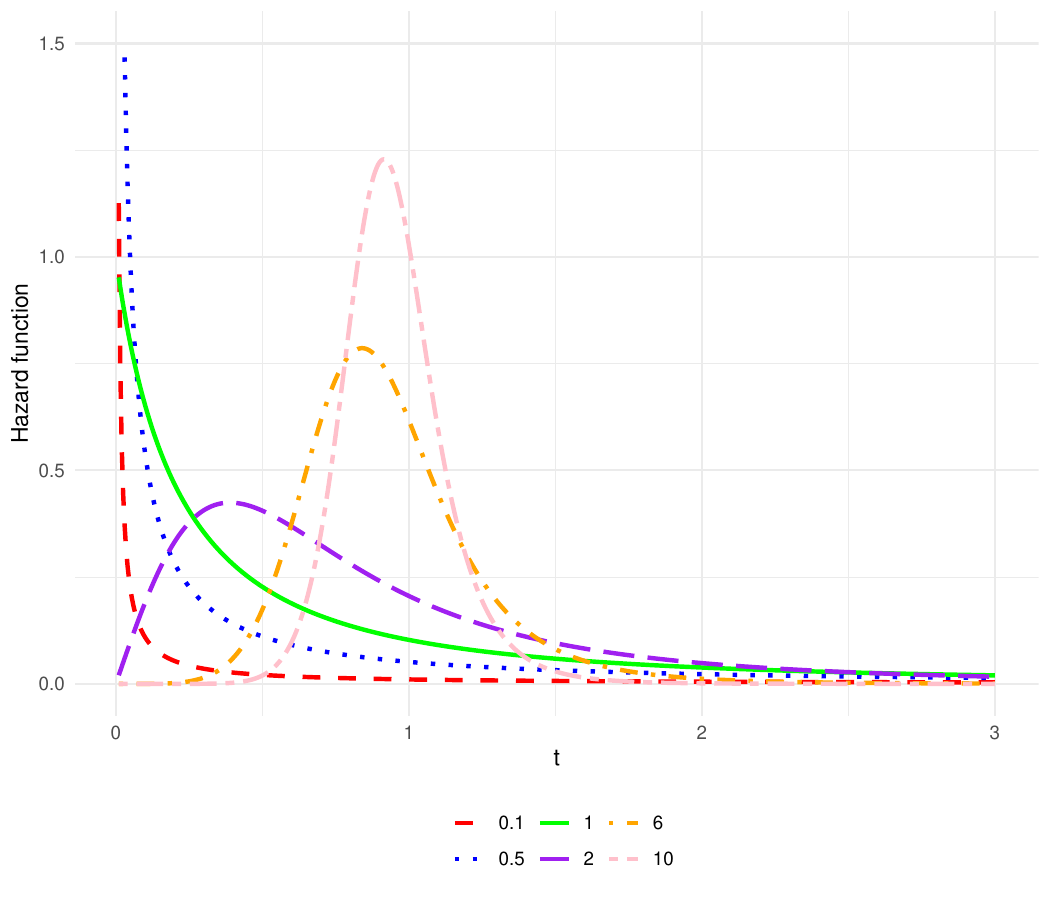} \\
		\caption{Hazard function of DD-PVF distribution for different values of $\alpha$ (lines colors and style).}
		\label{fig_HF_DD-PVF}
	\end{center}
\end{figure}

The hazard function of the DD-PVF distribution is given by
\begin{equation*}
\label{dd_pvf_hazard}
\displaystyle h(t\mid \mbox{\boldmath{$\vartheta$}})=\frac{\alpha\beta\theta^2t^{-(\alpha+1)}}{(\beta + \theta t ^{-\alpha})(\beta + \theta t^{-\alpha}-\theta\beta)}\left[1-\frac{\sigma^2}{1-\gamma}\log\left(\frac{\beta + \theta t^{-\alpha}-\theta\beta}{\beta + \theta t^{-\alpha}}\right)\right]^{\gamma-1}.
\end{equation*}
 The shape of hazard function of the DD-PVF distribution is decreasing for $\alpha\leq 1$ and a unimodal function for $\alpha > 1$ (Figure \ref{fig_HF_DD-PVF}).

The cure fraction of the DD-PVF frailty model is given by
\begin{eqnarray}
\label{p0_dagum_pvf}
\displaystyle \lim_{t\rightarrow\infty}S(t\mid \mbox{\boldmath{$\vartheta$}})=\exp\left(\frac{1-\gamma}{\gamma \sigma^2}\left\{1-\left[1-\frac{\sigma^2}{1-\gamma}\log\left(1-\theta\right)\right]^{\gamma}\right\}\right) = p_0(\theta,\gamma,\sigma^2) = p_0.
\end{eqnarray} 

If $\theta=1$, $p_0=0$ in the Equation \eqref{p0_dagum_pvf}, thus there is no cure rate according to the proposed model. %, and \eqref{surv_dagum_pvf} and \eqref{dens_dagum_pvf} are proper survival and density functions. 

In order to model the cure rate $p_0$ as a function of the covariate, the DD-PVF frailty distribution is reparameterized in terms of the cure fraction. By considering the relation presented in Equation \eqref{p0_dagum_pvf}, the $\theta$ parameter can be written as
\begin{eqnarray}
\label{eq-theta}
\theta= 1 - \exp\left(\frac{1-\gamma}{\sigma^2}\left\{1-\left[1-\frac{\gamma\sigma^2}{1-\gamma}\log(p_0)\right]^{\frac{1}{\gamma}}\right\}\right).\end{eqnarray}

The proposed approach allows us to estimate the cure fraction without including an additional parameter in the modeling, which is necessary for the mixture model approach \cite{boag1949maximum}. Besides, if $\theta=1$ in the proposed model, it indicates there is no cure. 

If we replace \eqref{eq-theta} in \eqref{surv_dagum_pvf} and \eqref{dens_dagum_pvf}, we have the reparameterized density and survival functions of the DD-PVF model, respectively. Hereafter, we shall use the notation $T\sim DD-PVF(\alpha,\beta, p_0,\gamma, \sigma^2)$, where  $\alpha>0$, $\beta>0$, $0<\gamma< 1$, $\sigma^2 > 0$ and $0\leq p_0 < 1$.  
%Thus, the proposed model under the generalized Gompertz distribution is referred to as Generalized Gompertz Cure Rate Quantile Regression (GGCRQR). 

Now, we build the DD-PVF frailty regression model, considering the following functional relations:
\begin{eqnarray*}
%\label{reg_mu}
\alpha(\boldsymbol{\zeta},\textbf{w})= \alpha =\exp(\textbf{w}^{\top}\boldsymbol{\zeta}), \nonumber \\
%\label{reg_mu}
\beta(\boldsymbol{\eta},\textbf{x})= \beta =\exp(-\textbf{x}^{\top}\boldsymbol{\eta}) \nonumber
\end{eqnarray*}
and
\begin{equation*}
%\label{reg_p0}
p_0(\boldsymbol{\nu},\textbf{z})= p_{0} =\frac{\exp(\textbf{z}^{\top}\boldsymbol{\nu})}{1+\exp(\textbf{z}^{\top}\boldsymbol{\nu})}, \nonumber
\end{equation*}
in which $\textbf{w}^{\top}=(1,w_{1},\ldots,w_{q})$, $\textbf{x}^{\top}=(1,x_{1},\ldots,x_{p})$ and $\textbf{z}^{\top}=(1,z_{1},\ldots,z_{r})$ are the vectors of covariates; and $\boldsymbol{\zeta}=(\zeta_{0},\zeta_{1},\ldots,\zeta_{q})^{\top}$, $\boldsymbol{\eta}=(\eta_{0},\eta_{1},\ldots,\eta_{p})^{\top}$ and $\boldsymbol{\nu}=(\nu_{0},\nu_{1},\ldots,\nu_{r})^{\top}$  are the unknown vectors of regression parameters to be estimated. Thus, $\boldsymbol{\vartheta}=(\boldsymbol{\zeta}^{\top}, \boldsymbol{\eta}^{\top}, \boldsymbol{\nu}^{\top}, \gamma,\sigma^2)^{\top}$ is the vector of the parameters. Additionally, we highlight that the vectors of the covariates $\textbf{w}$, $\textbf{x}$, and $\textbf{z}$, as often happens in practice, can be the same.

It is worth mentioning that by considering regressors in the $\alpha$ parameter we can have different shapes for the hazard function by depending on values of $\textbf{w}$, as one can also observe in Figure \ref{fig_HF_DD-PVF}.

In short, some advantages of the DD-PVF frailty model:
\begin{itemize}
\item Possibility of estimating the cure fraction without adding an extra parameter dedicated to that, unlike the standard mixture model approach;
\item Evaluate the effect of covariates in the cure fraction directly, what it is not common in the defective distribution scenario;
\item Indicates that there is no cure fraction for a subgroup if $\theta$ parameter related to that is one;
\item Estimate different hazard functions depending on the covariate set, by considering regressors in $\alpha$;
\item Possibility of modeling unimodal hazard function. 
\end{itemize}

\section{Inference} \label{inference}

Consider that the lifetime $T$ is possibly not observed and it is subject to right-censored failure time. Let $C$ denote the censoring time, and a sample of size $n$, we then observe the $i$-th lifetime $t_{i}=\min\{T_i,C_i\}$, and $i$-th failure indicator $\delta_i=I(T_i\leq C_i)$, where $\delta_i=1$ if $T_i$ is observed and $\delta_i=0$ otherwise, for $i=1,\ldots,n$. 
 
We consider that $T_i$'s are independent random variables with density and survival functions  $S \left( t_i\mid \boldsymbol{\vartheta} \right)$ and $f \left( t_i\mid \boldsymbol{\vartheta}\right)$, obtained by replacing $\alpha_{i}=\exp(\textbf{w}^{\top}_i\boldsymbol{\zeta})$, $\beta_i= \exp(-\textbf{x}^{\top}_i\boldsymbol{\eta})$ and $p_{0i} =  \exp(\textbf{z}_i^{\top}\boldsymbol{\nu})[1+\exp(\textbf{z}_i^{\top}\boldsymbol{\nu})]^{-1}$ in Equations  \eqref{surv_dagum_pvf} and \eqref{dens_dagum_pvf}, respectively, where $\textbf{w}^{\top}_i=(1,w_{i1},\ldots,w_{iq})$, $\textbf{x}^{\top}_i=(1,x_{i1},\ldots,x_{ip})$ and $\textbf{z}^{\top}_i=(1,z_{i1},\ldots,z_{ir})$, for $i=1,\ldots,n$. We assume that each censoring time $C_i$ is independent of lifetime $T_i$, for all $i=1,\ldots,n$, and we consider a noninformative censoring assumption, i.e., the censoring distribution does not involve the parameters of the distribution of $T$.  
Therefore,  the likelihood function of $\boldsymbol{\vartheta}$ can be written as
\begin{equation}
\label{eq-9}
L(\boldsymbol{\vartheta};{\bf D})\propto\prod_{i=1}^{n}\left[f \left( t_i\mid \boldsymbol{\vartheta} \right)\right]^{\delta_i}\left[S \left( t_i\mid \boldsymbol{\vartheta} \right)\right]^{1-\delta_i},
\end{equation}
where ${\bf D} = \left(n, {\bf t}, {\boldsymbol \delta}, {\bf W},{\bf X}, {\bf Z} \right)$, with ${\bf t} = \left(t_1, \ldots, t_n \right)^{\top}, {\boldsymbol \delta} = \left(\delta_1, \ldots, \delta_n \right)^{\top}$.  

Furthermore, ${\bf W} = \left({\bf w}^{\top}_1, \ldots, {\bf w}^{\top}_n \right)$, ${\bf X} = \left({\bf x}^{\top}_1, \ldots, {\bf x}^{\top}_n \right)$ and ${\bf Z} = \left({\bf z}^{\top}_1, \ldots, {\bf z}^{\top}_n\right)$  are $n\times (q+1)$, $n\times (p+1)$ and $n\times (r+1)$ matrices containing the covariates information, respectively.

Maximum likelihood (ML) estimates for parameters from the DD-PVF frailty model are obtained by numerically maximizing the logarithm of the likelihood function \eqref{eq-9}, that is, $l(\boldsymbol{\vartheta};{\bf D})=\log L(\boldsymbol{\vartheta};{\bf D})$. Many routines are
available for numerical maximization. We used the {\tt optim} routine in the R software \cite{R1} for considering the Broyden–
Fletcher–Goldfarb–Shanno (BFGS) nonlinear optimization algorithm. 

Let $\widehat{\boldsymbol{\vartheta}}$ be the ML estimator of $\boldsymbol{\vartheta}$. Under certain standard regularity conditions, $\widehat{\boldsymbol{\vartheta}}$  is consistent and asymptotically follows a multivariate normal distribution with mean $\boldsymbol{\vartheta}$ and covariance matrix $\boldsymbol{\Sigma}(\widehat{\boldsymbol{\vartheta}})$. For large values of $n$, $\boldsymbol{\Sigma}(\widehat{\boldsymbol{\vartheta}}) = \boldsymbol{\mathcal{I}}^{-1}(\widehat{\boldsymbol{\vartheta}})$, where $\boldsymbol{\mathcal{I}}^{-1}(\widehat{\boldsymbol{\vartheta}})$
is the inverse expected Fisher information
matrix. However, this quantity is difficult to compute and we approximate it by the inverse observed Fisher information
matrix, denoted by $\boldsymbol{\mathcal{H}}^{-1}(\widehat{\boldsymbol{\vartheta}})$. Thus, approximate $(1-\gamma_1)100\%$ confidence intervals (CI) for each parameter are obtained. In order to estimate the proportion of long-term survivors and $\theta$ values, the invariance property of the ML estimators is considered, while the corresponding standard error can be estimated using the delta method \citep{lehmann2006theory}. 

Another approach can also be considered by identifying that $\gamma$ is a perturbation parameter and the maximum profile likelihood approach can be used to estimate the interest parameters $(\boldsymbol{\zeta}, \boldsymbol{\eta}, \boldsymbol{\nu}, \sigma^2)$ \citep{severini2000likelihood}.  The estimation process is done in two steps: the first one is dedicated to choosing the best value for $\gamma$, say $\gamma_{*}$, and the second one is dedicated to obtaining the maximum profile likelihood estimates of $\boldsymbol{\vartheta}_{-\gamma}$, which indicates the vector $\boldsymbol{\vartheta}$ without the $\gamma$ parameter, by considering the $\gamma_{*}$ value, that is, the maximum of the logarithm of the profile likelihood $L(\boldsymbol{\vartheta}_{-\gamma}, \gamma_{*};{\bf D})$. The $\gamma_{*}$ can be obtained from a grid of values of $\gamma$ that maximizes the profile likelihood function.

In the next section, simulation studies are performed to investigate if the usual properties of the ML estimators hold, once it is not trivial to verify it analytically for the proposed model. We also evaluate the performance of the model under different values of $\sigma^2$ and compare the proposed model to its version without the frailty term. Besides, by the simulation studies, we investigate the sensitivity of the proposed model in identifying the existence or not of long-term survivors in a subgroup.  

\section{Simulation studies} \label{simulation}
 Monte Carlo simulation studies were conducted to evaluate the frequentist properties of the estimators of the proposed model and to compare its results with the DD model (model without frailty) results. There are two scenarios, each one based on one of the applications considered in this article.

\subsection{Scenario 1}
In this scenario, only a binary predictor variable is considered, denoted by $x$, and its values were generated from a Bernoulli distribution with parameter $0.5$. The true parameter values used in the data generation are based on the estimates obtained in the maternal population application (Subsection \ref{application_maternal}), by considering only the group variable (pregnant or postpartum woman) as inspiration for the parameter values. 

Specifically, the following parameter values were used: $\zeta_0 = 0.7$, $\zeta_1 = -0.16$ (obtaining $\alpha = \exp(0.7) = 2.01$, for $x= 0$, and $\alpha = \exp(0.7 - 0.16) = 1.72$, for $x= 1$), $\eta_0 = 7.3$ and $\eta_1 = -1.4$ (obtaining $\beta = \exp(-7.3) = 0.0006$, for $x= 0$, and $\beta = \exp(-7.3 + 1.4) = 0.0027$, for $x= 1$), $\nu_0 = -0.6$ and $\nu_1 = -1.5$, resulting in $p_0 = 0.35$, when $x=0$, and $p_0= 0.11$, when $x=1$. Different sample sizes are considered $n = 50,~ 100,~ 200, ~ 400, ~ 800,~ 1000,~2000,~5000,~8000$ and $10000$ and different values for $\sigma^2$, $\sigma^2 = 0, ~ 0.05, ~ 0.1, ~ 0.3, ~0.5, ~ 0.7, ~ 1, ~1.3,~1.5 $ and $1.7$. The $\gamma$ parameter is fixed $\gamma \rightarrow 0$ (DD-Gamma model) in order to corroborate the results obtained in the application (Subsection \ref{application_maternal}).

For each combination of parameter values and sample size, $1000$ datasets are generated. To introduce random censoring, a uniform distribution within the interval \((0, \tau)\) was assumed for the censoring times. The value of \(\tau\) was chosen to control the proportion of right-censored observations. The observed times and censoring indicators were generated according to the steps of the algorithm \ref{generation}.

\begin{algorithm}[h!]
    \caption{Data generation algorithm for the scenario 1.}\label{generation}
	\begin{algorithmic}[1]
	\State Determine desired values for $\zeta_0$, $\zeta_1$, $\eta_0$, $\eta_1$, $\nu_0$ and $\nu_0$;
	\State Define the proportion of censored data, given by $pc$;
	\State For the $i$th subject, draw $x_i\sim$ Bernoulli($0.5$), and calculate $p_{0x_i}$, for $x_i=0$ and $x_i=1$; 
	\State Draw $u_i\sim \mbox{Uniform}(0, 1)$. If $u_i< p_{0x_i}$, set $w_i = \infty$; otherwise, generate $u_{1i}\sim U(0,1-p_{0x_i})$ and calculate
	\begin{eqnarray*}
	   w_i&=&\left(\frac{\exp(-\eta_0-\eta_1x_i)\left\{\exp\left[\frac{1-(1-u_{1i})^{-\sigma^2}}{\sigma^2}\right] + \theta_i - 1\right\}}{\theta_i\left\{1-\exp\left[\frac{1-(1-u_{1i})^{-\sigma^2}}{\sigma^2}\right]\right\}}\right)^{-\frac{1}{\exp(\zeta_0+\zeta_1x_i)}}, ~ \mbox{with} ~ x_i=1 ~ \mbox{or} ~ x_i=0 ,~\mbox{and} \\
     \theta_i &=& 1 - \exp\left( \frac{1 - p_{0x_i}^{-\sigma^2}}{\sigma^2}\right);
	\end{eqnarray*}
    \State Draw $c_i\sim U(0,\tau_i)$, where $\tau_i$ is defined to have approximately $pc$ proportion of censoring data;
	\State Determine $t_i=\min\{w_i,c_i\}$. If $t_i=w_i$, set $\delta_i=1$, otherwise $\delta_i=0$;
	\State Repeat steps 3 to 6 for all $i=1,\ldots,n$.  The data set for the $i$th subject is $\{t_i,x_i, \delta_i\}, \ i=1, \ldots, n$.
	\end{algorithmic}
\end{algorithm}

\begin{figure}[h!]
\begin{minipage}[t]{0.5\linewidth}
  \centering
  \includegraphics[width=\linewidth]{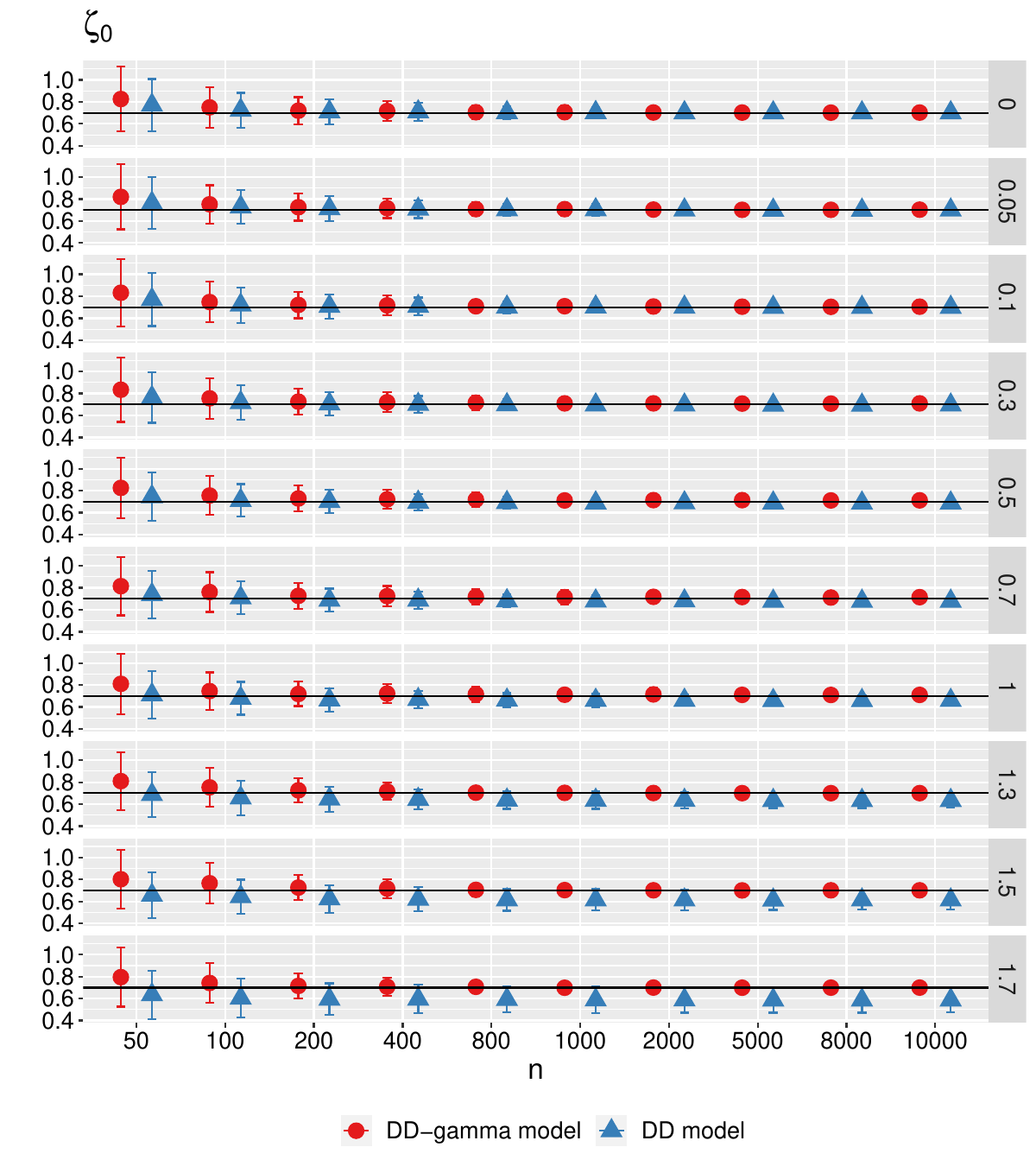}
\end{minipage}%
\begin{minipage}[t]{0.5\linewidth}
  \centering
  \includegraphics[width=\linewidth]{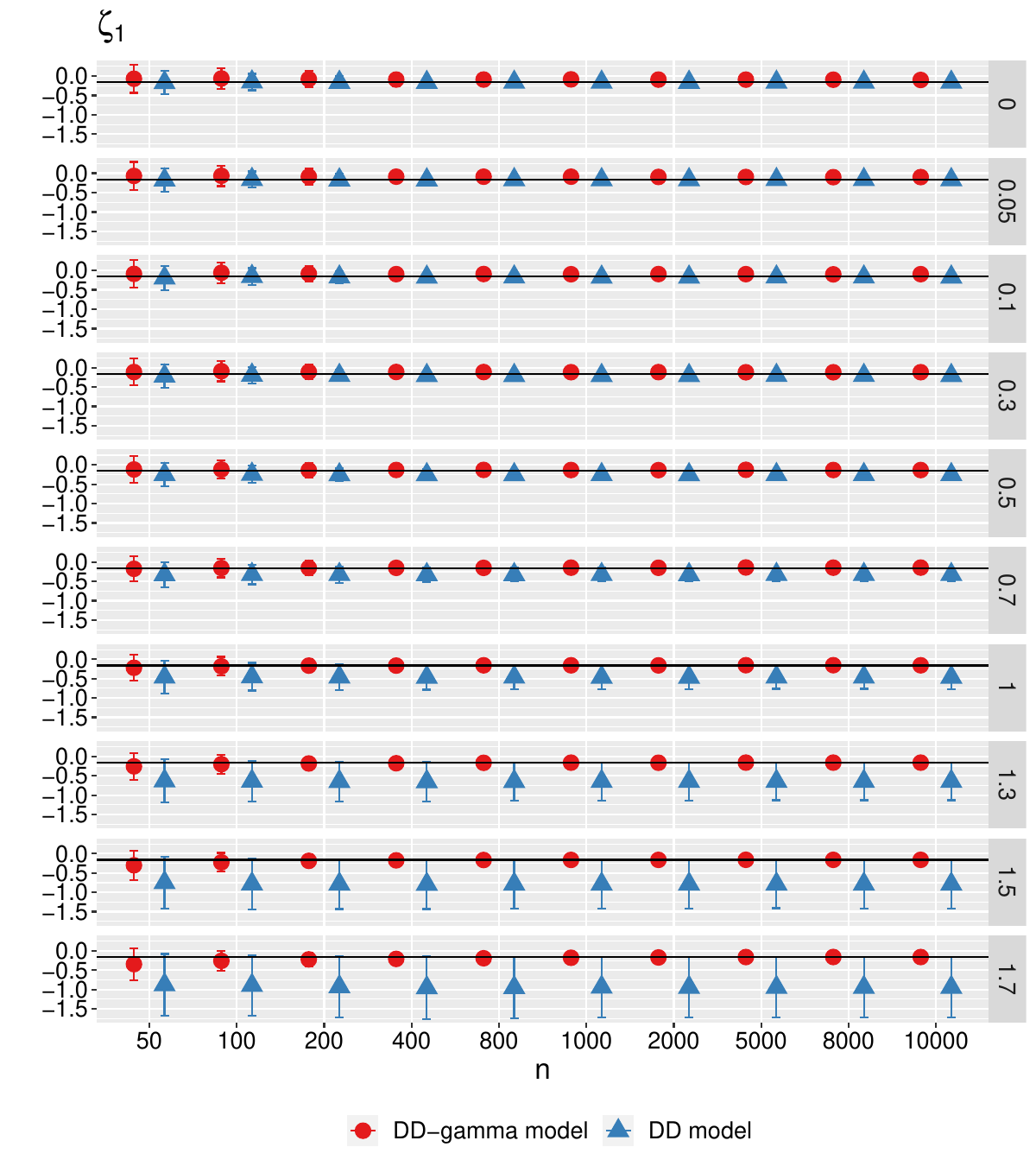}
\end{minipage}
\caption{Bias (symbols) and root mean square error
(bars) for $\zeta_0$ (left) and $\zeta_1$ (right) for each $\sigma^2$ value (block) for scenario 1.}
\label{scenario1_zeta}
\end{figure}

\begin{figure}[h!]
\begin{minipage}[t]{0.5\linewidth}
  \centering
  \includegraphics[width=\linewidth]{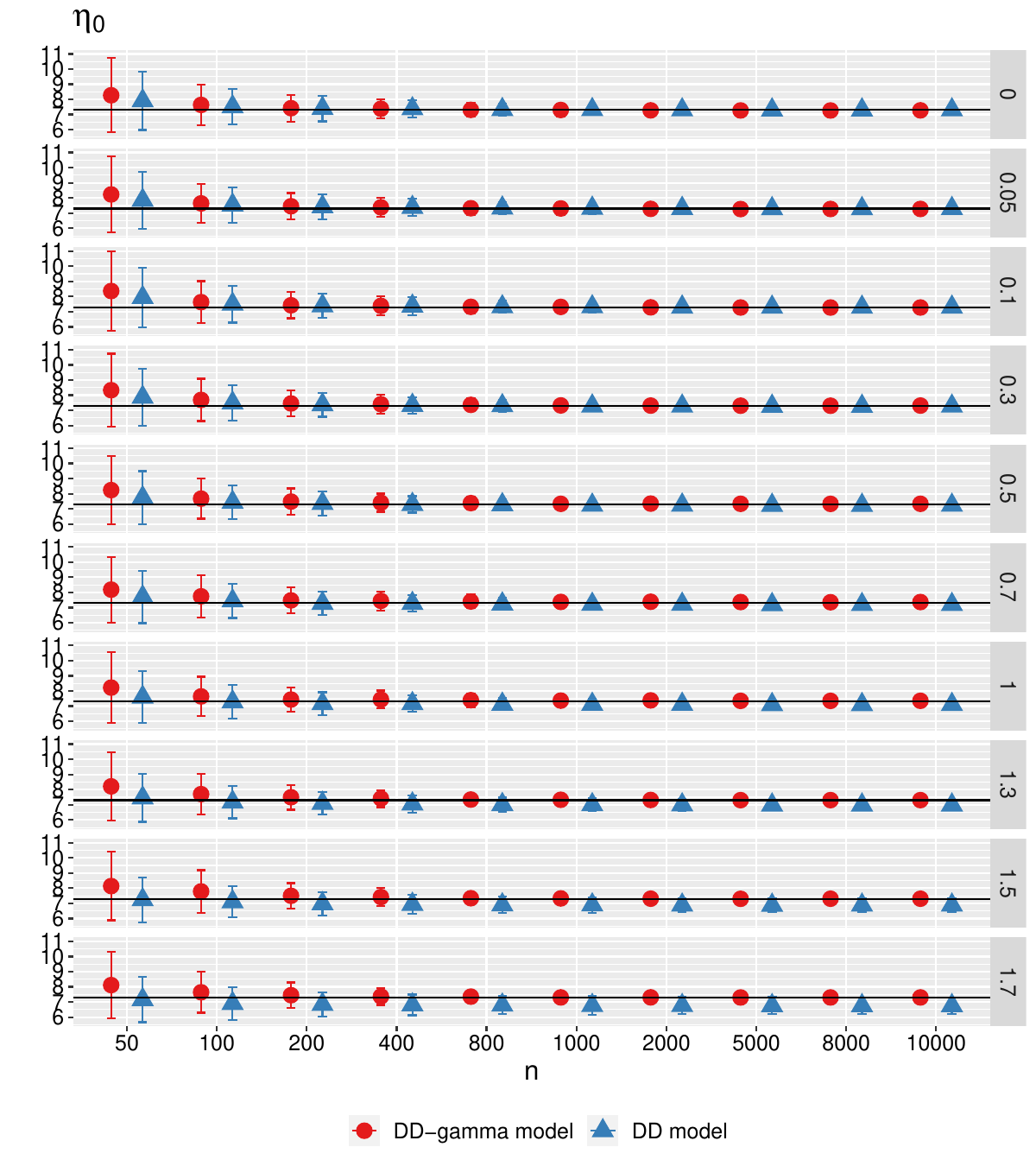}
\end{minipage}%
\begin{minipage}[t]{0.5\linewidth}
  \centering
  \includegraphics[width=\linewidth]{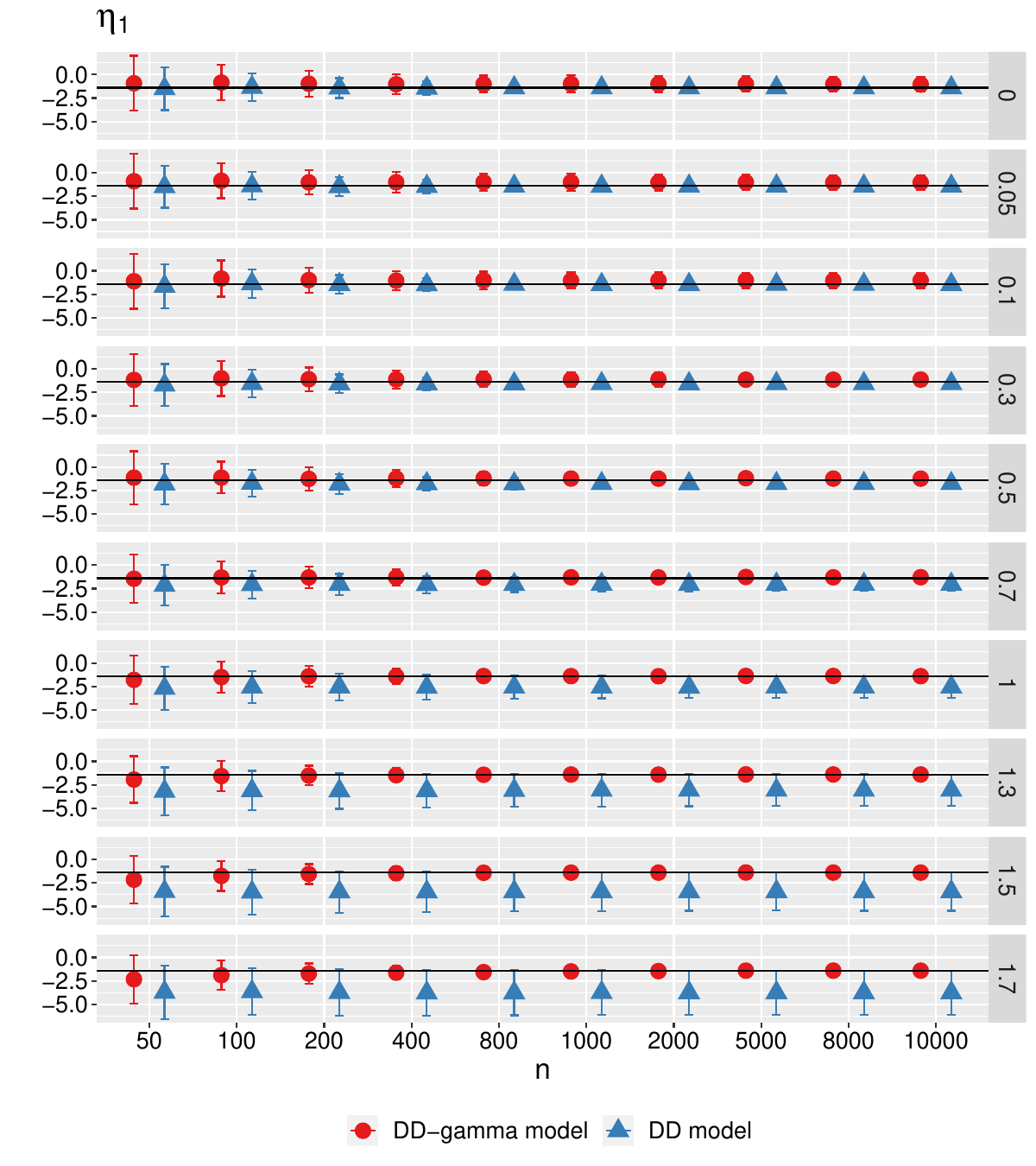}
\end{minipage}
\caption{Bias (symbols) and root mean square error
(bars) for $\eta_0$ (left) and $\eta_1$ (right) for each $\sigma^2$ value (block) for scenario 1.}
\label{scenario1_eta}
\end{figure}

\begin{figure}[h!]
\begin{minipage}[t]{0.5\linewidth}
  \centering
  \includegraphics[width=\linewidth]{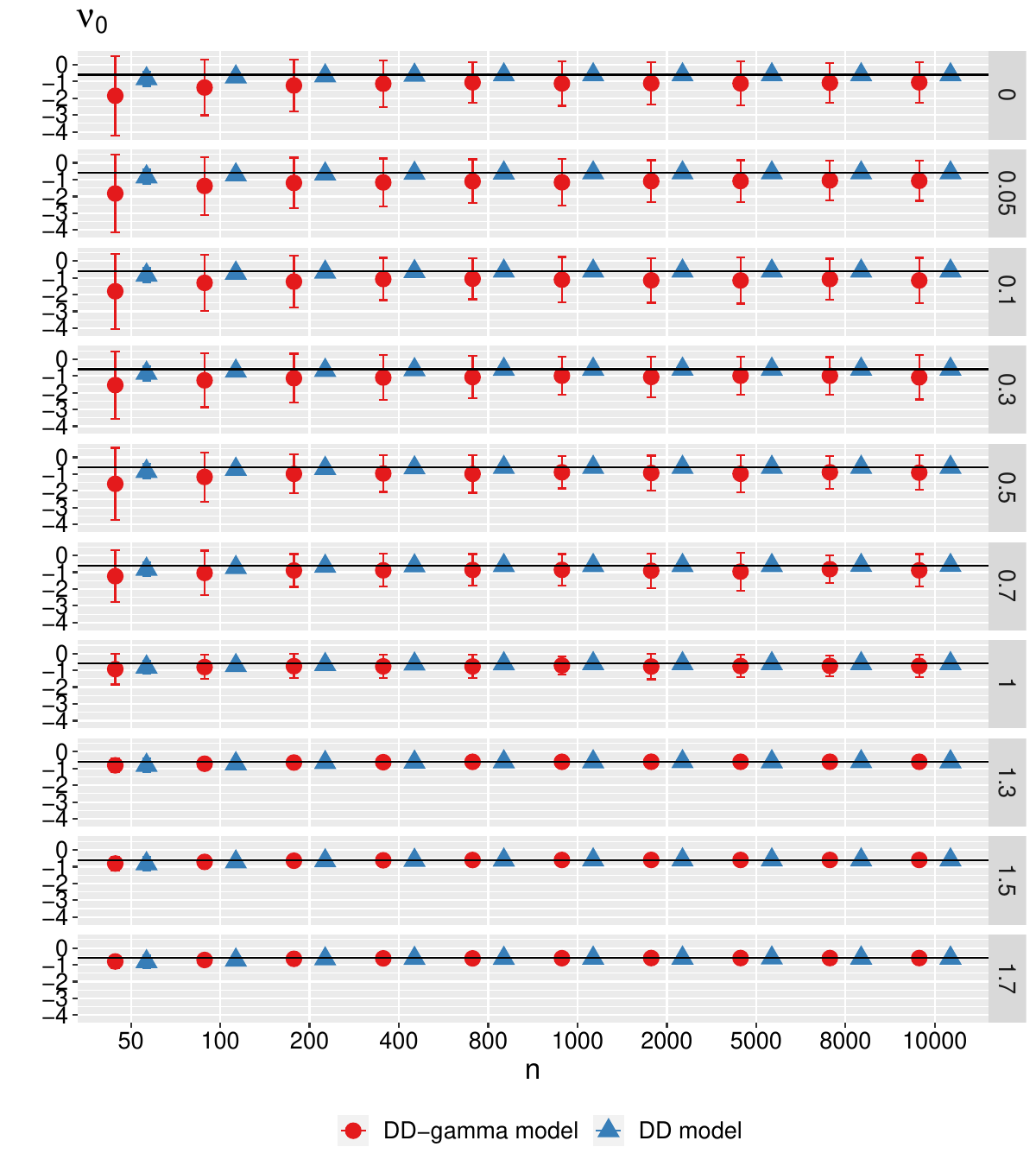}
\end{minipage}%
\begin{minipage}[t]{0.5\linewidth}
  \centering
  \includegraphics[width=\linewidth]{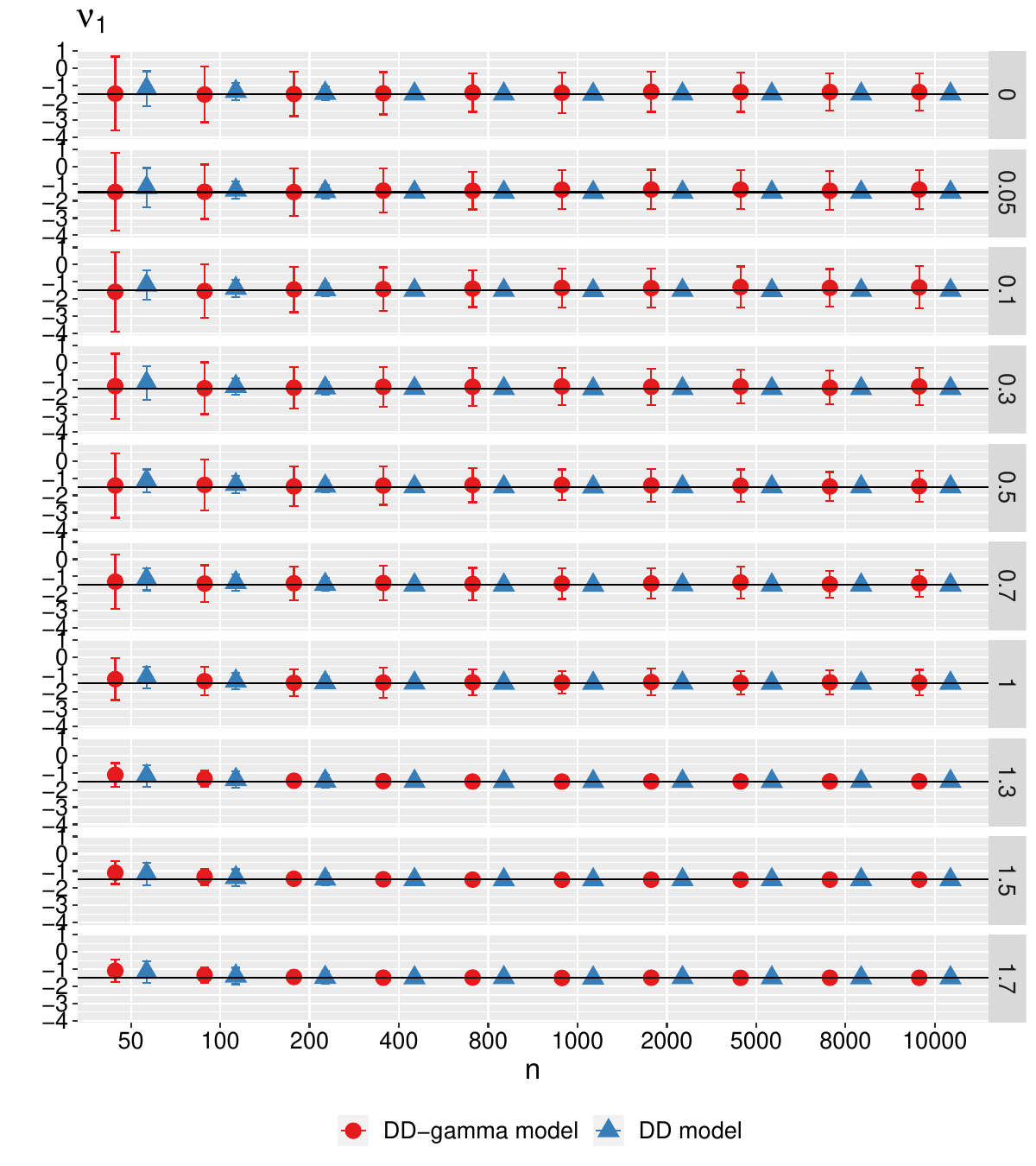}
\end{minipage}
\caption{Bias (symbols) and root mean square error
(bars) for $\nu_0$ (left) and $\nu_1$ (right) for each $\sigma^2$ value (block) for scenario 1.}
\label{scenario1_nu}
\end{figure}

In figures \ref{scenario1_zeta}, \ref{scenario1_eta} and \ref{scenario1_nu} are presented the bias and the root mean square error
(RMSE) of the MLEs of the parameters. Regarding the DD-gamma model, it can be observed that RMSE values decrease as the sample size increases, as well as bias, approaching zero. It's worth noting that the proposed model exhibits remarkable performance even in a scenario without frailty ($\sigma^2=0$). The performance of the DD model gets worse as the $\sigma^2$ value increases, mainly for the $\zeta_1$ and $\eta_1$ parameters.

We investigated the performance of many information criteria in the selection of the correct model between DD-gamma and DD models. We consider the Akaike information criterion (AIC), corrected Akaike information criterion (AICc), Bayesian information criterion (BIC), Hannan–Quinn information criterion (HQIC), and consistent Akaike information criterion (CAIC). They are given by: $AIC = -2l + 2k$, $AICc = AIC +2k(k+1)/(n-k-1)$, $BIC = -2l + k \log(n)$, $HQIC = -2l + 2k\log[\log(n)]$ and $CAIC = -2l + k[\log(n)+1]$, in which $l$ is the maximized log-likelihood function value, $k$ is the number of parameters in the fitted model, and $n$ is the sample size.

\begin{figure}[h!]
\centering
  \includegraphics[width=0.5\linewidth]{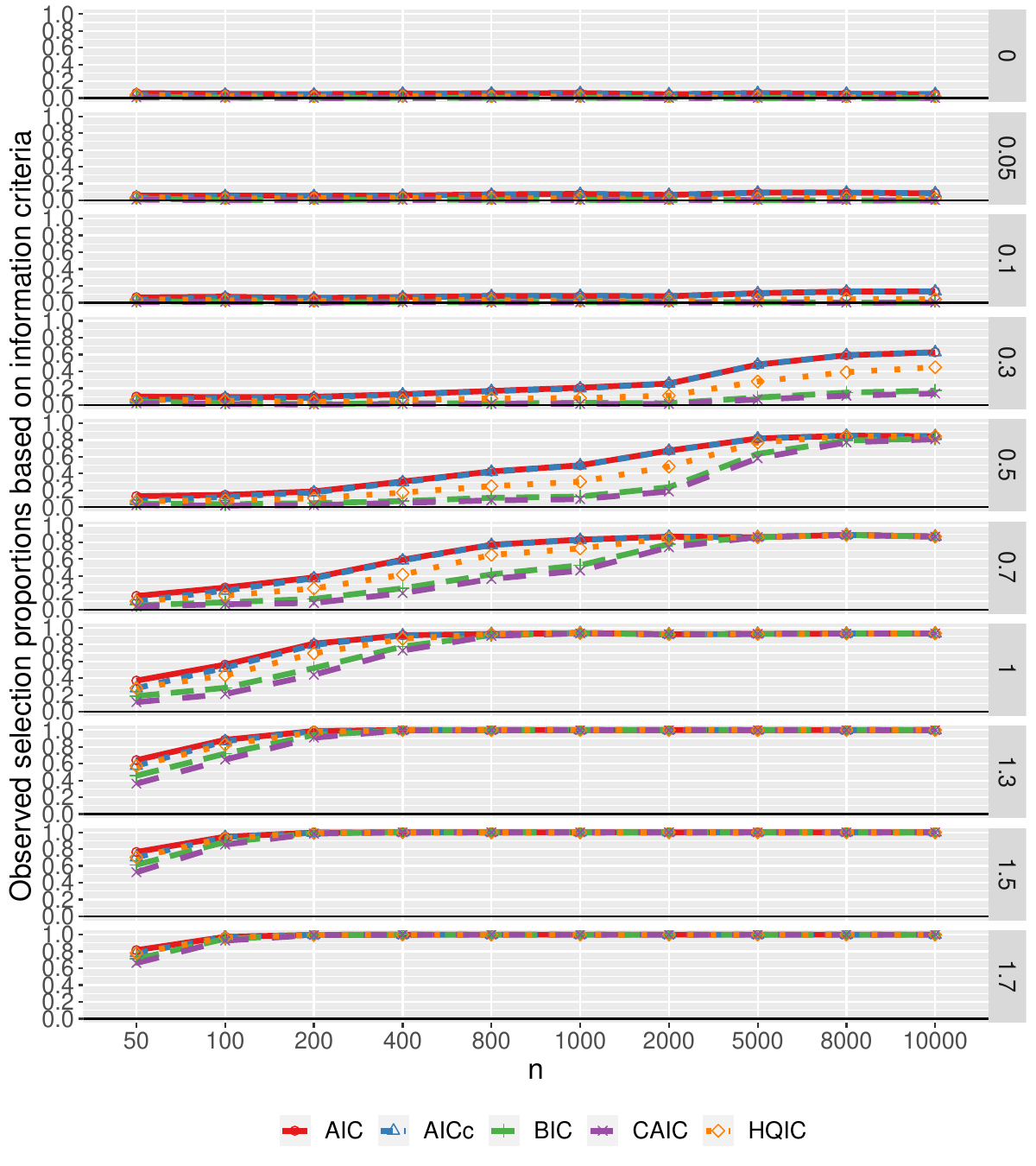}
\caption{Proportion of cases out of 1000 datasets where the DD-gamma model has better information criteria than the DD model for all $\sigma^2$ values (blocks) for scenario 1.}
\label{scenario1_porc}
\end{figure}

Figure \ref{scenario1_porc} shows the observed selection proportions in which the DD-gamma model is preferred for each of the five criteria and for all combinations of sample size and $\sigma^2$ values. We observe that the percentage choice of the DD-gamma model is close to zero for all information criteria when $\sigma^2$ is close to zero. Values of $\sigma^2$ greater than $0.3$ the percentage choice of the DD-gamma model is greater than 50\%, approaching 100\% as $\sigma^2$ gets larger.  These results show that the information criteria can distinguish between the models in the presence of high heterogeneity. For $\sigma^2 = 0.5$ and $0.7$, intermediary values of $\sigma^2$, the proportions for the correct models are higher for AIC and AICc, and thus we will use the AIC in Section \ref{application} as the model selection criteria.

\subsection{Scenario 2}

This second simulation scenario is based on the malignant neoplasms of skin (not melanoma) application (Subsection \ref{application_fosp}), in which there is one binary predictor and one group has a cured fraction while the other one does not have long-term survivors. 

As in the application, fortunately, only a small fraction of the patients is in the metastasis stage, we consider $x$ generated from a Bernoulli distribution with parameter $0.1$. Besides, the following parameter values were used: $\zeta_0 = 0.45$, $\zeta_1 = 0.64$ (obtaining $\alpha = \exp(0.45) = 1.56$, for $x= 0$, and $\alpha = \exp(0.45 + 0.64) = 2.97$, for $x= 1$), $\eta_0 = 8$ and $\eta_1 = -1.5$ (obtaining $\beta = \exp(-8) = 0.0003$, for $x= 0$, and $\beta = \exp(-8 + 1.5) = 0.0015$, for $x= 1$), $\nu_0 = 1.5$ and $\nu_1 = -5$, resulting in $p_0 = 0.81$, when $x=0$, and $p_0= 0.02$, when $x=1$. Different sample sizes are considered $n = 100,~ 200, ~ 400, ~ 800,~ 1000,~2000,~5000,~8000, ~ 10000, ~12000 $ and $15000$ and two values of $\sigma^2$ are considered: $\sigma^2=5$ and $\sigma^2=11$ (this one is the estimated value of $\sigma^2$ in the malignant neoplasms of skin application). The DD-PVF model is considered in order to corroborate the results obtained in the application (Subsection \ref{application_fosp}), and the $\gamma$ value is $0.73$. Based on the parameter values, by considering the Equation \eqref{eq-theta},  the $\theta$ values are: for $\sigma^2 = 5$, $\theta = 0.24$ when $x=0$, and $\theta = 1$ when $x=1$, and for $\sigma^2 = 11$, $\theta = 0.28$ when $x=0$ and $\theta = 1$ when $x=1$.

For each combination of parameter values and sample size, $1000$ datasets are generated. To introduce random censoring, a uniform distribution within the interval \((0, \tau)\) was assumed for the censoring times. The value of \(\tau\) was chosen to control the proportion of right-censored observations. The observed times and censoring indicators were generated according to the steps of the algorithm \ref{generation2}.

\begin{algorithm}[!h]
    \caption{Data generation algorithm for the scenario 2.}\label{generation2}
	\begin{algorithmic}[1]
	\State Determine desired values for $\zeta_0$, $\zeta_1$, $\eta_0$, $\eta_1$, $\nu_0$, $\nu_0$, $\gamma$ and $\sigma^2$;
	\State Define the proportion of censored data, given by $pc$;
	\State For the $i$th subject, draw $x_i\sim$ Bernoulli($0.1$), and calculate $p_{0x_i}$, for $x_i=0$ and $x_i=1$; 
        \State Calculate 
     \begin{eqnarray*}
     \theta_i = 1 - \exp\left( \frac{1-\gamma}{\sigma^2}\left\{1-\left[1-\frac{\gamma\sigma^2}{1-\gamma}\log(p_{0x_i})\right]^{\frac{1}{\gamma}}\right\}
 \right);
	\end{eqnarray*}
    \State Draw $u_i\sim \mbox{Uniform}(0, 1)$. If $\theta_i = 1$, calculate
      \begin{eqnarray*}
	   w_i&=&\left[\frac{\exp(-\eta_0-\eta_1x_i)\exp\left(\frac{1-\gamma}{\sigma^2}\left\{1-\left[1-\log(1-u_{i})\frac{\gamma\sigma^2}{1-\gamma}\right]^{-\frac{1}{\gamma}}\right\} \right)}{1-\exp\left(\frac{1-\gamma}{\sigma^2}\left\{1-\left[1-\log(1-u_{1i})\frac{\gamma\sigma^2}{1-\gamma}\right]^{-\frac{1}{\gamma}}\right\} \right)}\right]^{-\frac{1}{\exp(\zeta_0+\zeta_1x_i)}}, ~ \mbox{with} ~ x_i=1 ~ \mbox{or} ~ x_i=0; 
	\end{eqnarray*}
    If $\theta_i < 1$ and $u_i< p_{0x_i}$, set $w_i = \infty$; otherwise, generate $u_{1i}\sim U(0,1-p_{0x_i})$ and calculate
	\begin{eqnarray*}
	   w_i&=&\left\{\frac{\exp(-\eta_0-\eta_1x_i)\left[\exp\left(\frac{1-\gamma}{\sigma^2}\left\{1-\left[1-\log(1-u_{1i})\frac{\gamma\sigma^2}{1-\gamma}\right]^{-\frac{1}{\gamma}}\right\} \right) + \theta_i - 1\right]}{\theta_i\left[1-\exp\left(\frac{1-\gamma}{\sigma^2}\left\{1-\left[1-\log(1-u_{1i})\frac{\gamma\sigma^2}{1-\gamma}\right]^{-\frac{1}{\gamma}}\right\} \right)\right]}\right\}^{-\frac{1}{\exp(\zeta_0+\zeta_1x_i)}}, ~ \mbox{with} ~ x_i=1 ~ \mbox{or} ~ x_i=0; 
	\end{eqnarray*}
    \State Draw $c_i\sim U(0,\tau_i)$, where $\tau_i$ is defined to have approximately $pc$ proportion of censoring data;
	\State Determine $t_i=\min\{w_i,c_i\}$. If $t_i=w_i$, set $\delta_i=1$, otherwise $\delta_i=0$;
	\State Repeat steps 3 to 7 for all $i=1,\ldots,n$.  The data set for the $i$th subject is $\{t_i,x_i, \delta_i\}, \ i=1, \ldots, n$.
	\end{algorithmic}
\end{algorithm}

\begin{figure}[h!]
\begin{minipage}[t]{0.5\linewidth}
  \centering
  \includegraphics[width=\linewidth]{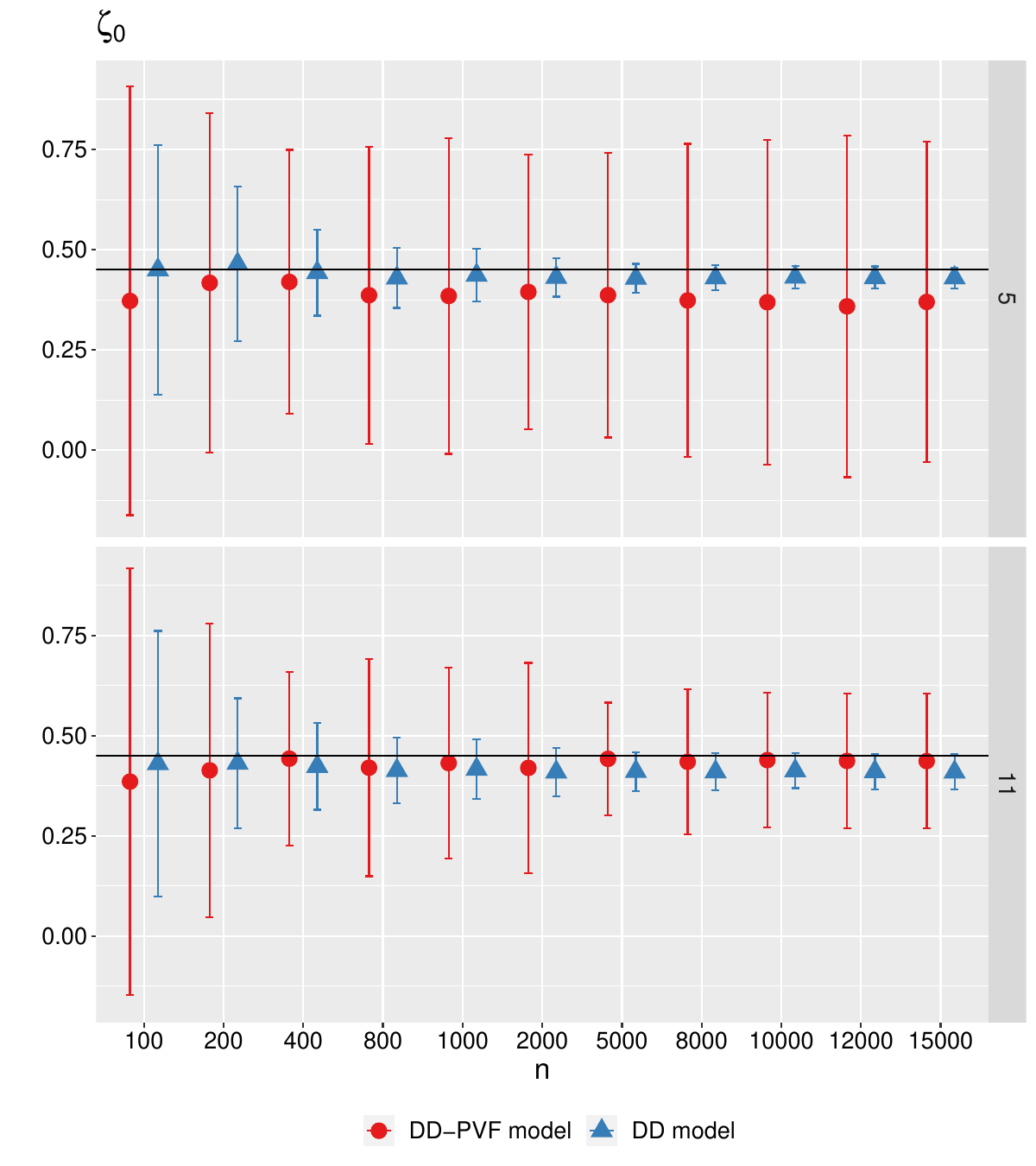}
\end{minipage}%
\begin{minipage}[t]{0.5\linewidth}
  \centering
  \includegraphics[width=\linewidth]{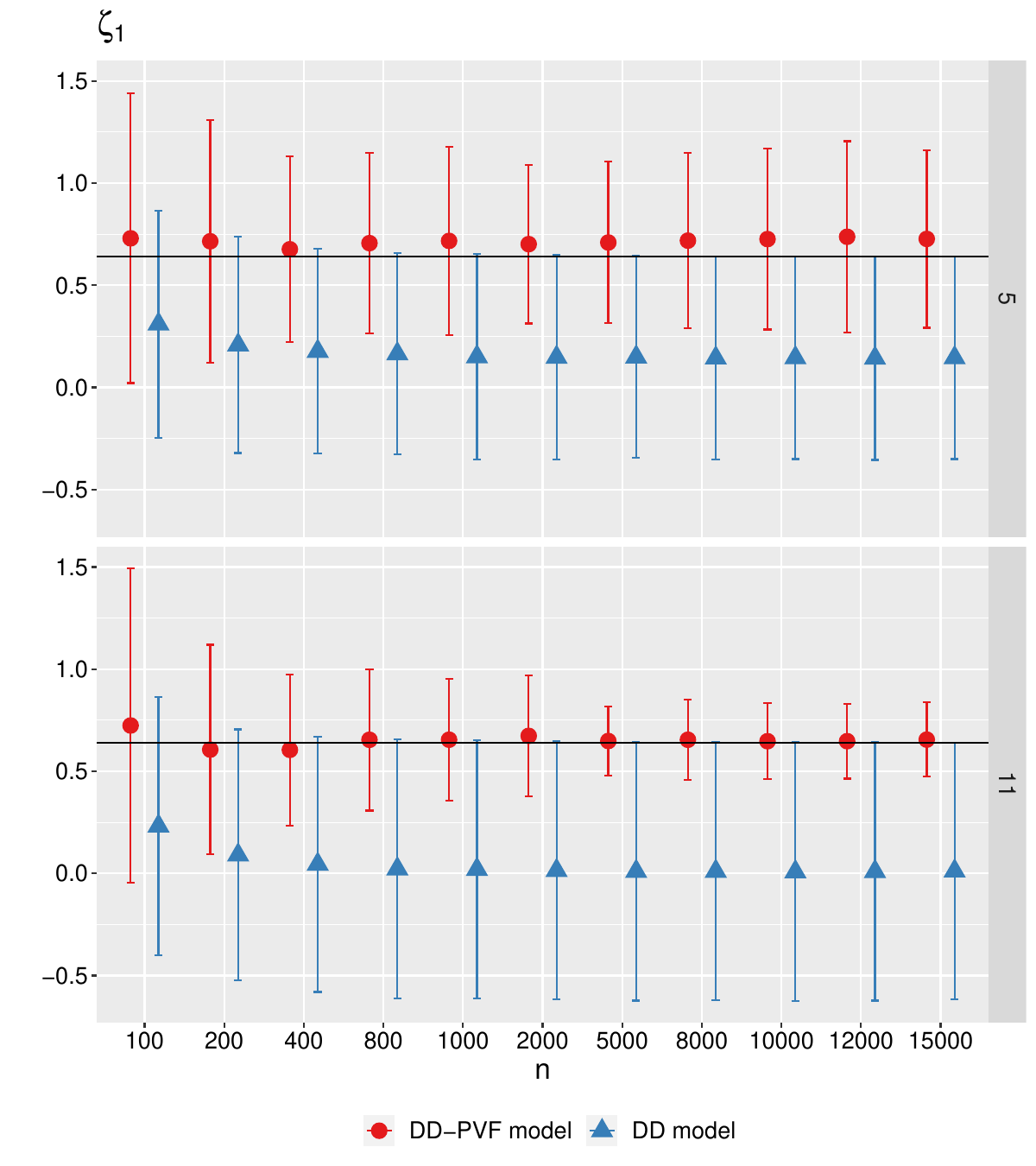}
\end{minipage}
\caption{Bias (symbols) and root mean square error
(bars) for $\zeta_0$ (left) and $\zeta_1$ (right) for each $\sigma^2$ value (block) for scenario 2.}
\label{scenario2_zeta}
\end{figure}

\begin{figure}[h!]
\begin{minipage}[t]{0.5\linewidth}
  \centering
  \includegraphics[width=\linewidth]{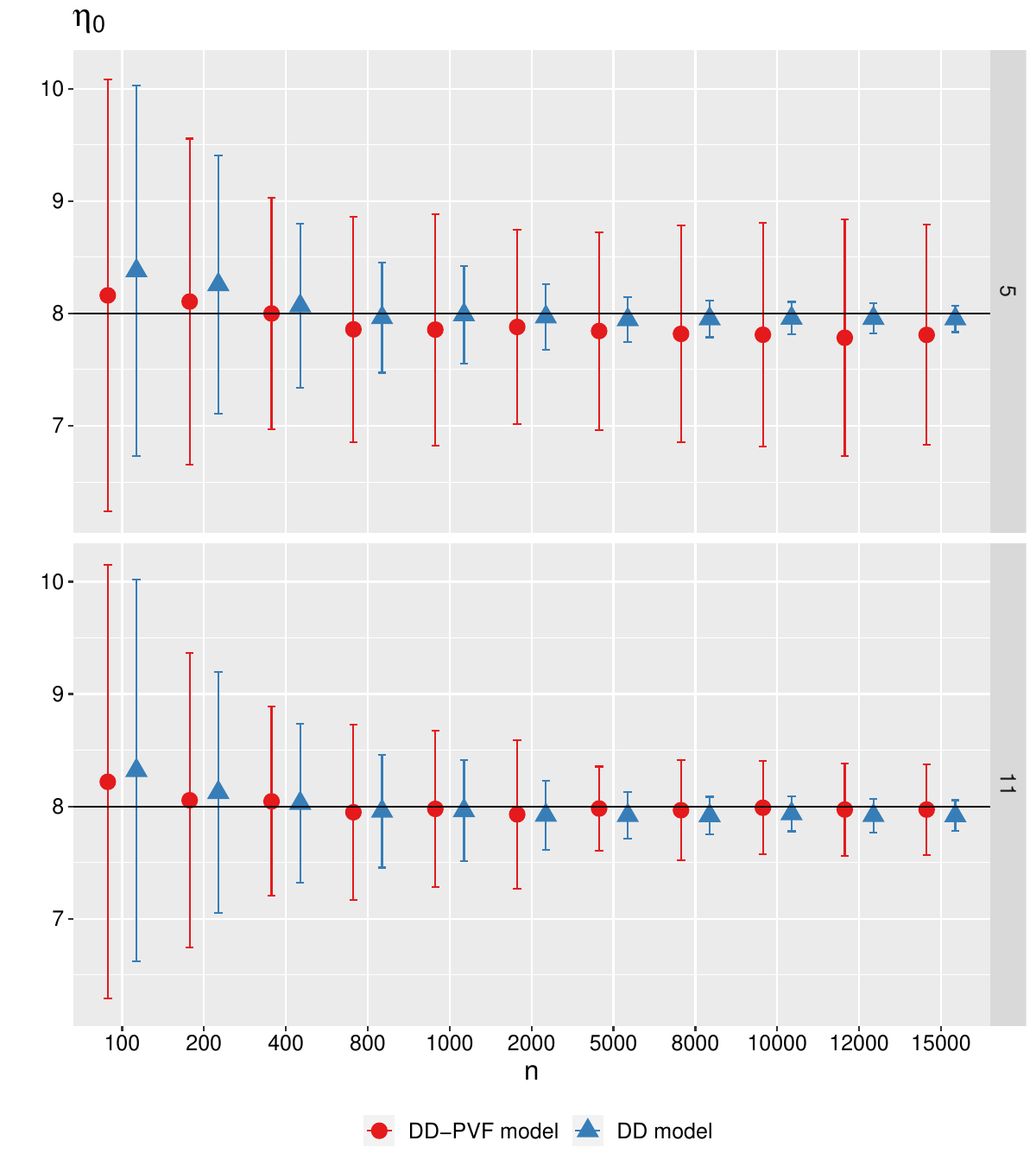}
\end{minipage}%
\begin{minipage}[t]{0.5\linewidth}
  \centering
  \includegraphics[width=\linewidth]{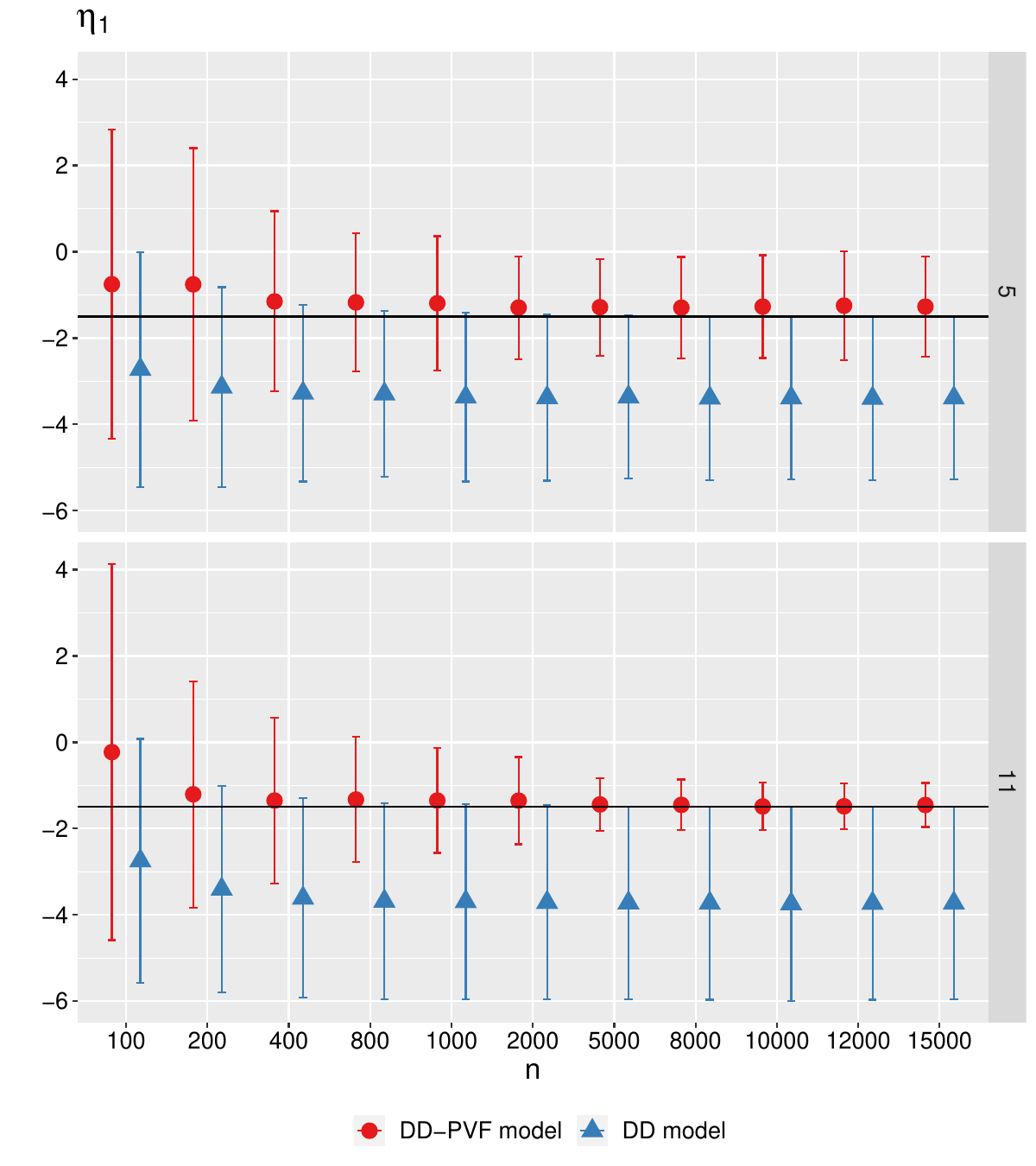}
\end{minipage}
\caption{Bias (symbols) and root mean square error
(bars) for $\eta_0$ (left) and $\eta_1$ (right) for each $\sigma^2$ value (block) for scenario 2.}
\label{scenario2_eta}
\end{figure}

\begin{figure}[h!]
\begin{minipage}[t]{0.5\linewidth}
  \centering
  \includegraphics[width=\linewidth]{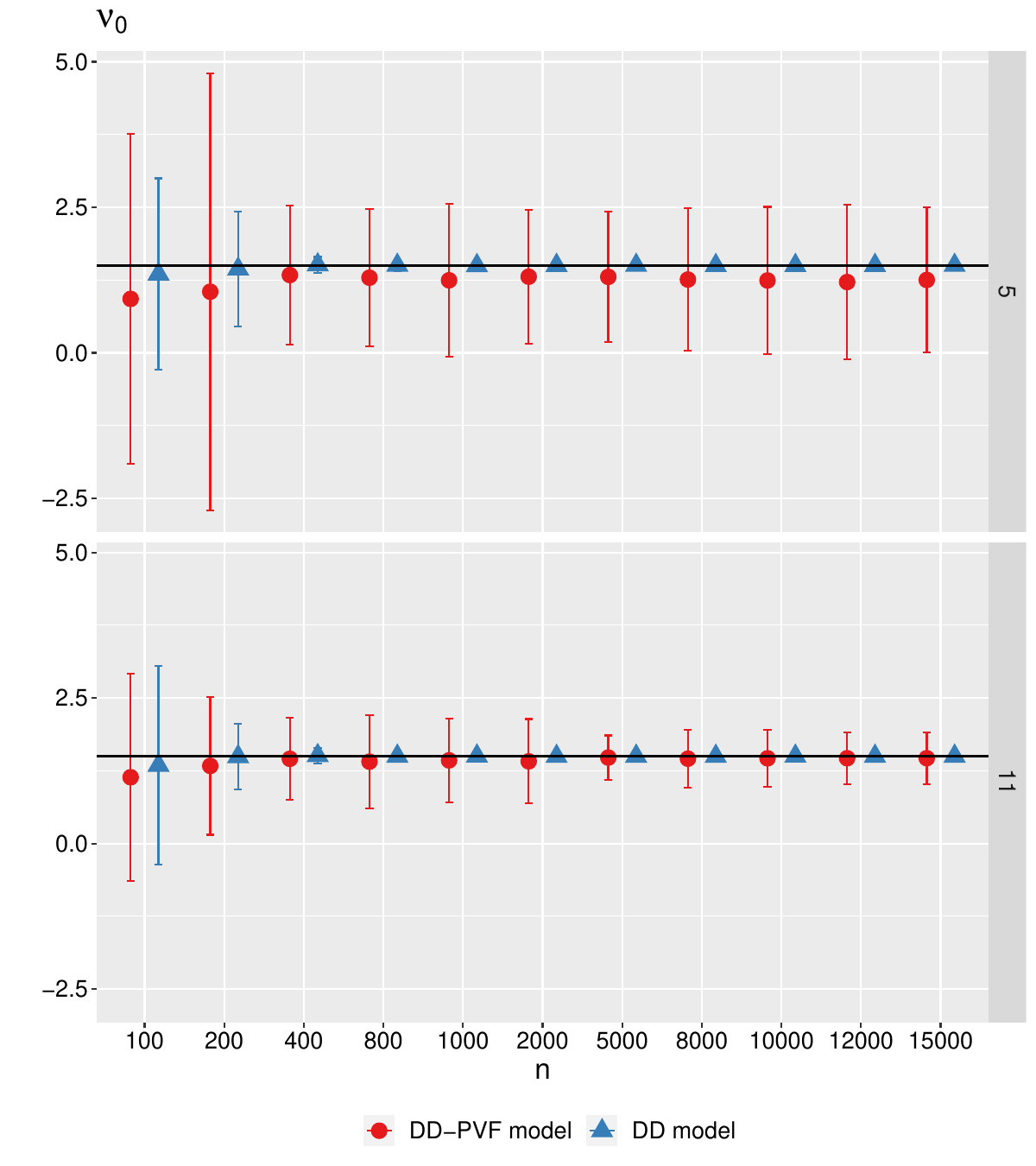}
\end{minipage}%
\begin{minipage}[t]{0.5\linewidth}
  \centering
  \includegraphics[width=\linewidth]{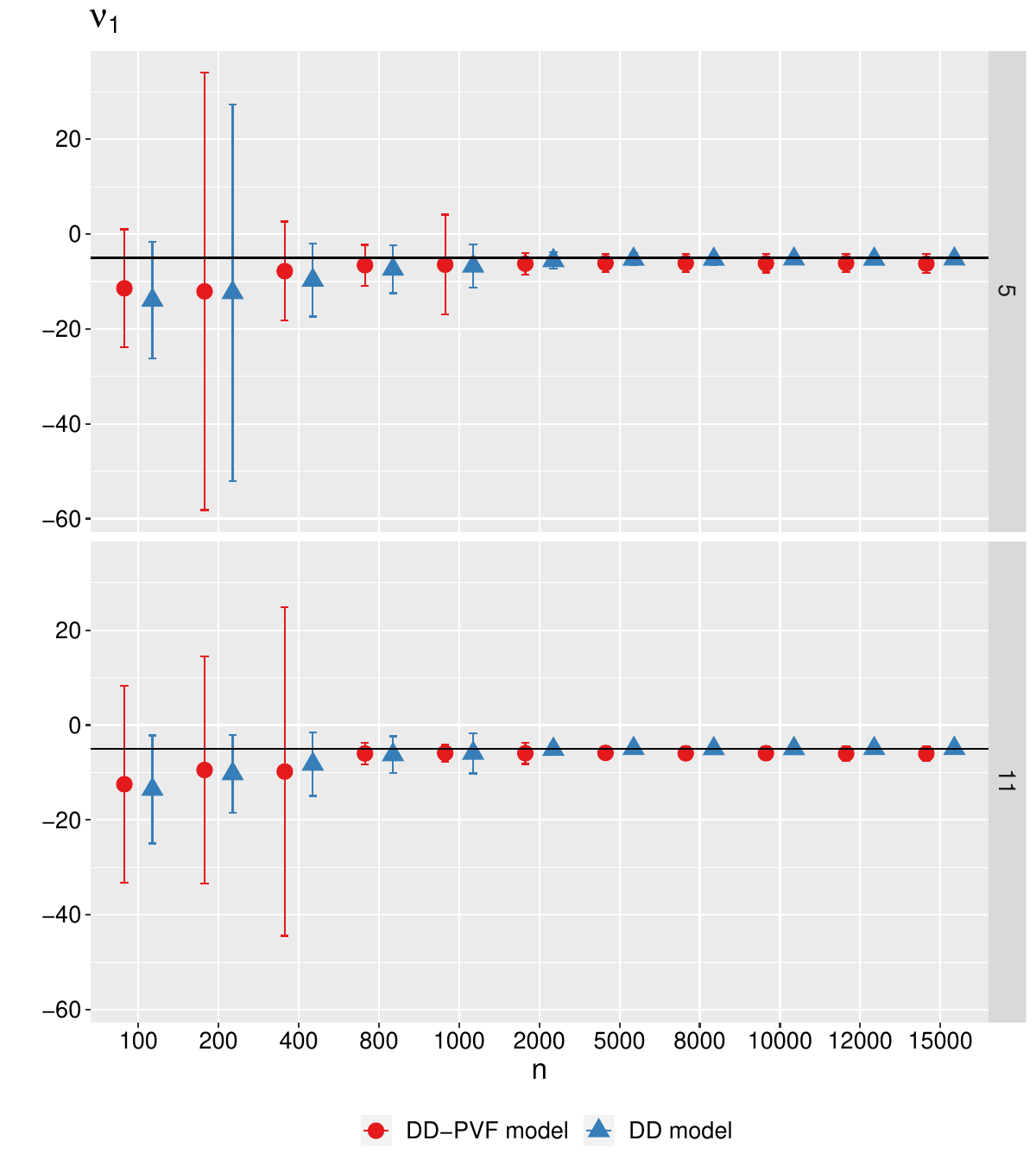}
\end{minipage}
\caption{Bias (symbols) and root mean square error
(bars) for $\nu_0$ (left) and $\nu_1$ (right) for each $\sigma^2$ value (block) for scenario 2.}
\label{scenario2_nu}
\end{figure}

\begin{figure}[h!]
\begin{minipage}[t]{0.5\linewidth}
  \centering
  \includegraphics[width=\linewidth]{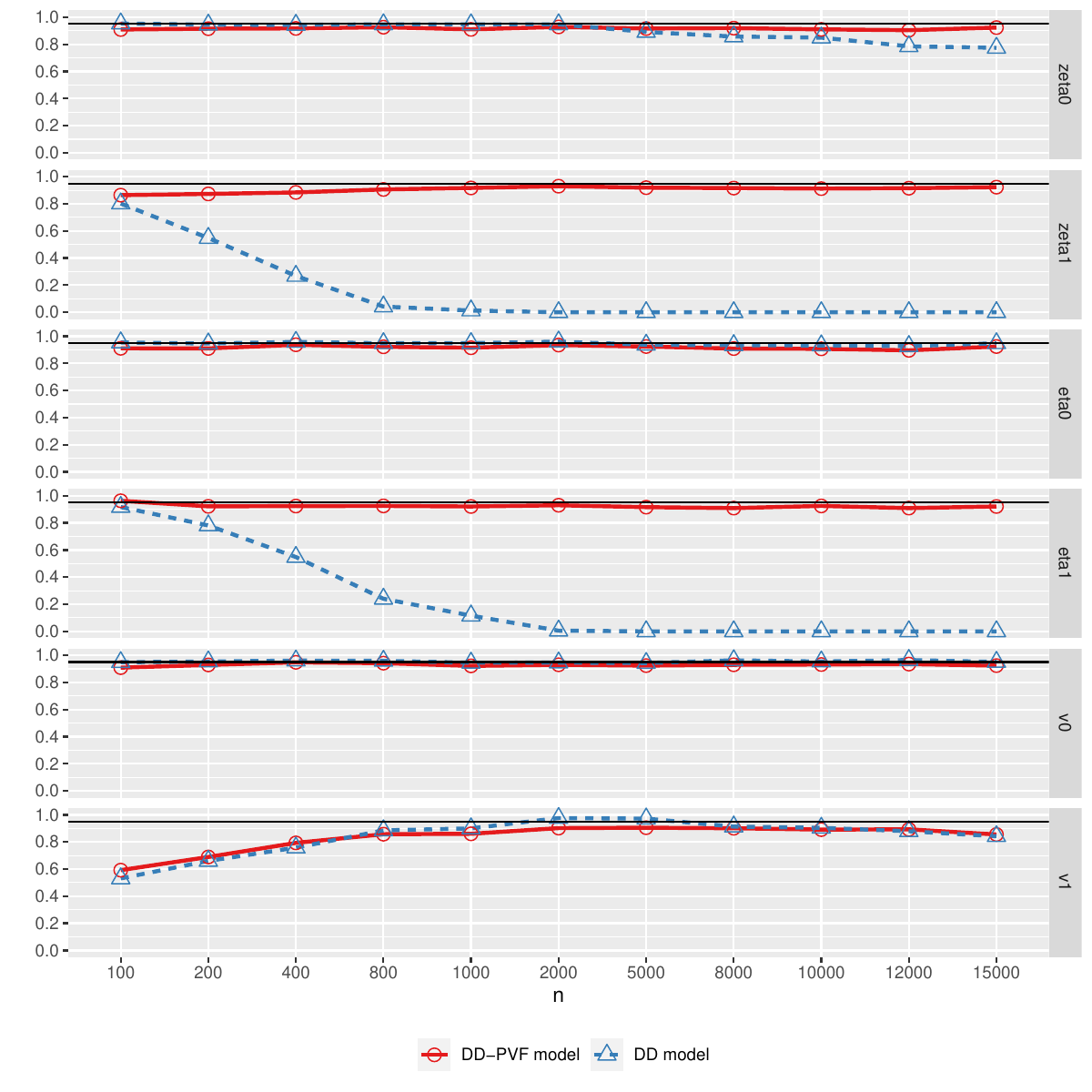}
\end{minipage}
\begin{minipage}[t]{0.5\linewidth}
  \centering
  \includegraphics[width=\linewidth]{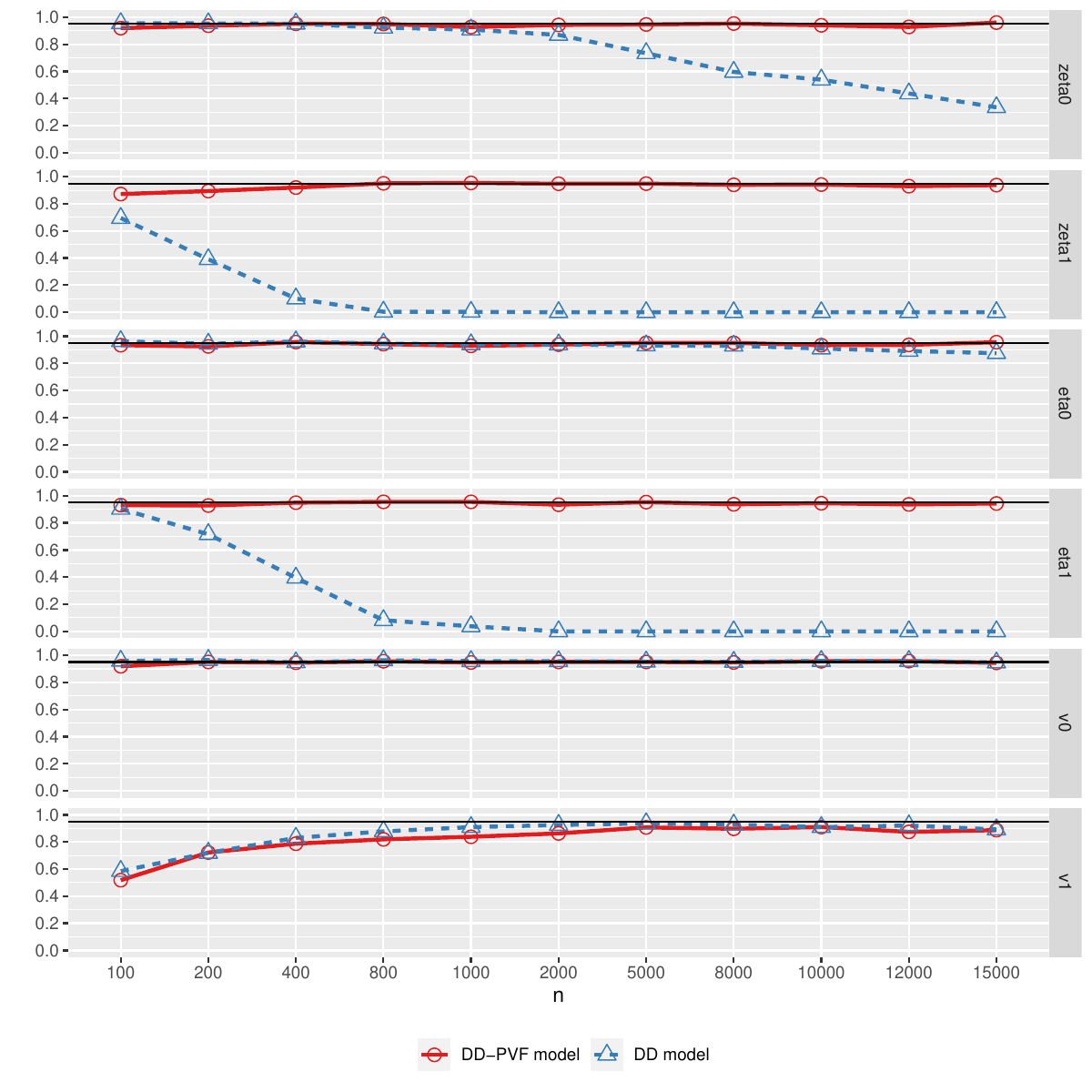}
\end{minipage}
\caption{Empirical coverage probability of 95\% confidence interval for $\sigma^2=5$ (in the left), and $\sigma^2=11$ (in the right) for scenario 2.}
\label{scenario2_coverage}
\end{figure}

The DD-PVF model's parameters were estimated via maximum likelihood and via maximum profile likelihood. The second approach presented a lower variance and, because of this, it is presented here.  
In figures \ref{scenario2_zeta}, \ref{scenario2_eta} and \ref{scenario2_nu} are presented the bias and the RMSE of the MLEs of the parameters. It can be observed that RMSE values decrease as the sample size increases, as well as bias, approaching zero for the DD-PVF model. The performance of the DD model is worse than the DD-PVF model's, mainly for the $\zeta_1$ and $\eta_1$ parameters, for all values of $\sigma^2$.  Figure \ref{scenario2_coverage} presents the 95\% coverage probability (CP) for DD-PVF and DD models for all parameters, where we can observe that, in general, the CP of DD-PVF remains close to the nominal level of 95\%, while the DD model deviates significantly from the expected level for  $\zeta_0$, $\zeta_1$, and $\eta_1$, mainly when $\sigma^2=11$. 

The proportions of cases in which 95\% confidence intervals contain 1 value for $\theta$ parameter are presented in tables \ref{proportion_theta_scenario2_sigma5} and \ref{proportion_theta_scenario2_sigma11}, for $\sigma^2=5$ and $\sigma^2=11$, respectively. Both models correctly resulted in proportions close to zero for $\theta$ when $x=0$, once its true value is $0.24$ (for $\sigma^2 = 5$) and $0.28$ (for $\sigma^2 = 11$). On the other hand, for $\theta$ when $x=1$, only DD-PVF model obtained high proportions, close to the nominal level of $95\%$ mainly for large sample sizes, while the DD model demonstrated a significantly underwhelming performance concerning the 95\% coverage probability.

\begin{table}[h!]
\caption{Proportion of cases in which 95\% confidence intervals contain 1 for $\theta$ parameter, for $\sigma^2=5$ in the scenario 2. \\ $\theta_0$ means $\theta$ when $x=0$ and $\theta_1$ means $\theta$ when $x=1$.}
\begin{tabular}{c|c|ccccccccccc}
\hline
  &  &  \multicolumn{11}{c}{$n$}\\
\hline
Parameter       & Model        &  100 & 200   & 400   & 800   & 1000  & 2000  & 5000  & 8000  & 10000    & 12000 & 15000 \\
\hline
\multirow{2}{*}{$\theta_0$} & DD model     & 0.001	& 0.002	&  0 &	0 &	0 &	0 &	0 &	0 &	0 &	0	& 0  \\
                        & DD-PVF model & 0.089	& 0.042 &	0.019 &	0.028 &	0.033 &	0.026 &	0.032 &	0.036	& 0.039 &	0.041 &	0.039
 \\
\hline
\multirow{2}{*}{$\theta_1$} & DD model     & 0.56	& 0.68 & 	0.766 & 	0.83 & 	0.793 & 	0.547 & 	0.055 & 	0.001 & 	0	&  0	&  0
    \\
                        & DD-PVF model & 0.73	& 0.797 &	0.862 &	0.927 &	0.935 &	0.954	& 0.969 &	0.963 &	0.965 &	0.967 &	0.962
 \\
\hline
\end{tabular}
\label{proportion_theta_scenario2_sigma5}
\end{table}

\begin{table}[h!]
\caption{Proportion of cases in which 95\% confidence intervals contain 1 for $\theta$ parameter, for $\sigma^2=11$ in the scenario 2. \\ $\theta_0$ means $\theta$ when $x=0$ and $\theta_1$ means $\theta$ when $x=1$.}
\begin{tabular}{c|c|ccccccccccc}
\hline
  &  &  \multicolumn{11}{c}{$n$}\\
\hline
Parameter       & Model        &  100 & 200   & 400   & 800   & 1000  & 2000  & 5000  & 8000  & 10000    & 12000 & 15000 \\
\hline
\multirow{2}{*}{$\theta_0$} & DD model     & 0.009   & 0     & 0     & 0     & 0     & 0     & 0     & 0     & 0        & 0     & 0     \\
                        & DD-PVF model & 0.094   & 0.046 & 0.011 & 0.013 & 0.009 & 0.013 & 0.004 & 0.007 & 0.004    & 0.005 & 0.006 \\
\hline
\multirow{2}{*}{$\theta_1$} & DD model     & 0.612   & 0.762 & 0.833 & 0.757 & 0.687 & 0.307 & 0.007 & 0     & 0        & 0     & 0     \\
                        & DD-PVF model & 0.694   & 0.832 & 0.869 & 0.897 & 0.9   & 0.911 & 0.953 & 0.946 & 0.949 & 0.935 & 0.934 \\
\hline
\end{tabular}
\label{proportion_theta_scenario2_sigma11}
\end{table}

\section{Application} \label{application}

In this section, the two applications of the real data in this study are presented: the dataset of COVID-19 in the maternal population (Section \ref{application_maternal}) and the dataset of malignant neoplasms of skin (Section \ref{application_fosp}). 

Both are anonymized datasets with no possibility of individual identification, according to Brazilian regulations of the National Research Ethics Commission (Comissão Nacional de Ética em Pesquisa–CONEP), this study does not require prior approval by the institutional ethics board.

The free software R \citep{R1} was also used for the statistical analysis described here.

\subsection{COVID-19 in the maternal population} \label{application_maternal}

In this application, we investigate the impact of prognostic factors on the survival of pregnant and postpartum women hospitalized with severe acute respiratory syndrome (SARS) confirmed by COVID-19. These are the covariates: age group ($<$30 and $\geq$ 30), obstetric moment (pregnant or puerperal women), COVID-19 vaccine (yes or no), saturation (yes or no), loss of taste (yes or no) and obesity (yes or no).

We considered the database of the SARS Epidemiological Surveillance Information System (SIVEP-Gripe), the official system for recording cases and deaths due to SARS in Brazil, made available by the Ministry of Health, on the openDataSUS portal (\url{https://opendatasus.saude.gov.br/group/dados-sobre-srag}). 

There are 9,936 records of pregnant and puerperal women aged 10 to 55 years hospitalized with SARS due to COVID-19, confirmed by RT-PCR test, dated from the beginning of the pandemic until March 2022. Death caused by complications of the COVID-19 disease (1,198 - 12\% of the total sample) was considered as the event, and all other cases were considered as right-censored failure time (8,738 - 88\% of the total sample). The time, in days, from hospitalization to death due to COVID-19 is the lifetime variable $T$.

When evaluating the characterization variables, a total of 8,084 (81.36\%) were pregnant and 1,852 (18.64\%) were postpartum. Of these women, 4,519 (45.48\%) were younger than 30 years old and 5,417 (54.52\%) were 30 years old or older. Finally, 9,225 (92.84\%) pregnant and puerperal women hospitalized with severe COVID-19 did not take the vaccine to combat COVID-19, while only 711 (7.16\%) received the vaccine. When evaluating the symptom variables, 4,094 (41.20\%) had desaturation and 1,215 (12.23\%) had a loss of taste. When assessing the comorbidity variable, 716 (7.21\%) were classified as obese.

\begin{figure}[h!]
	\centering
	\begin{center}
		\includegraphics[scale = 0.75]{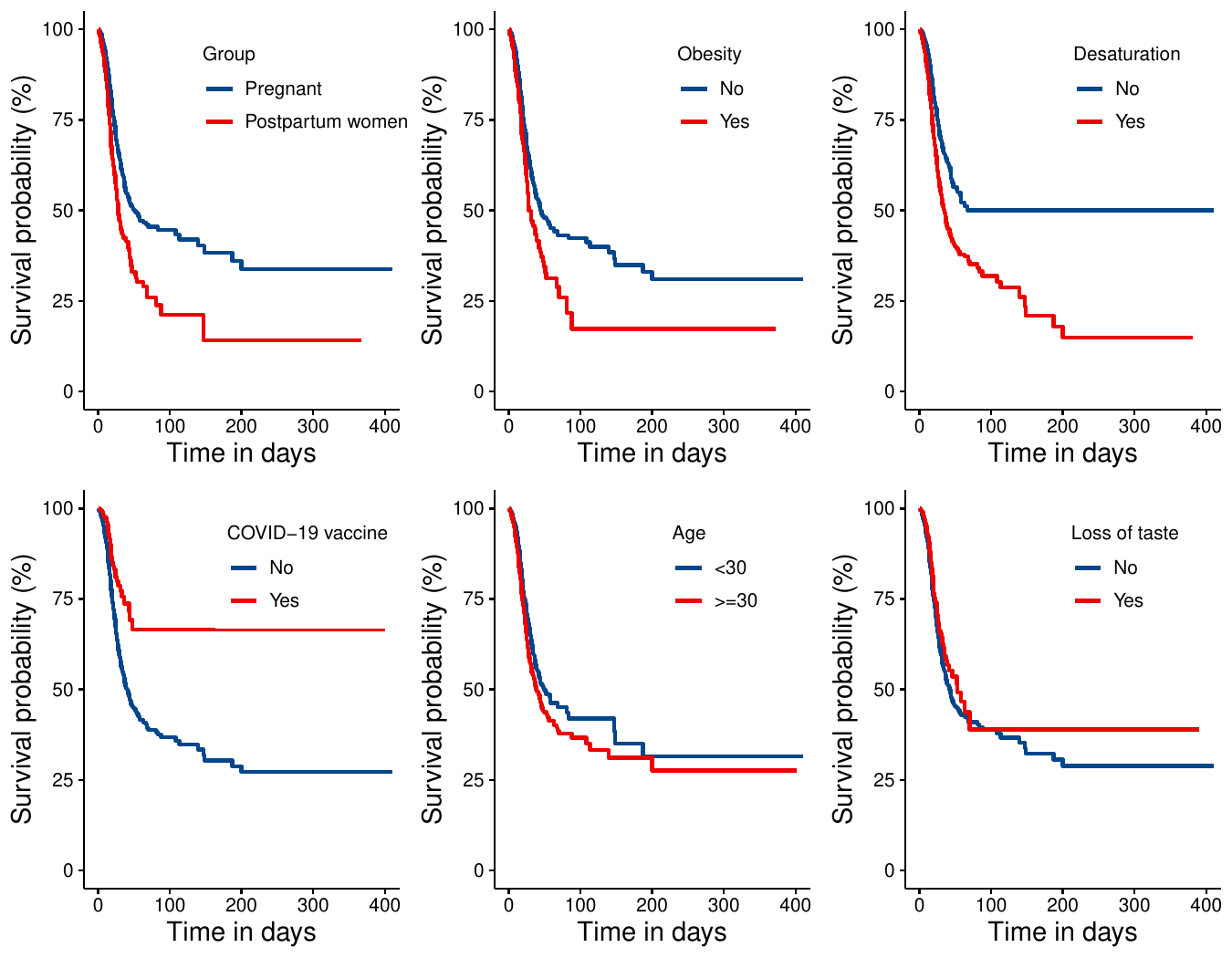} \\
		\caption{Kaplan Meier estimates for survival curves of pregnant and puerperal women hospitalized with SARS due to COVID-19.}
		\label{fig_KM}
	\end{center}
\end{figure}

Figure \ref{fig_KM} shows the Kaplan-Meier (K-M) estimates of the variables mentioned above. 
At first glance, we can see that a fraction of the population of women under investigation was cured, and therefore, using an approach that considers the cure is indeed appropriate for the problem in question. Despite this, we note that the COVID-19 pandemic seems to have been more lethal in women in the puerperal cycle and in women over 30 years of age. On the other hand, it can be seen that those who were vaccinated against the virus received the best prognosis.

In addition, the presence of some symptoms can serve as an indication for the worsening of the disease, since they can reduce the probability of survival of the hospitalized patient. These are the cases of symptoms related to respiratory failure, such as saturation. On the other hand, there are symptoms that are associated with mild cases of COVID-19, such as loss of taste \cite{lechien2021prevalence}, very different from the situation in which pregnant and postpartum women are obese.

\begin{figure}[h!]
	\centering
	\begin{center}
		\includegraphics[scale = 0.75]{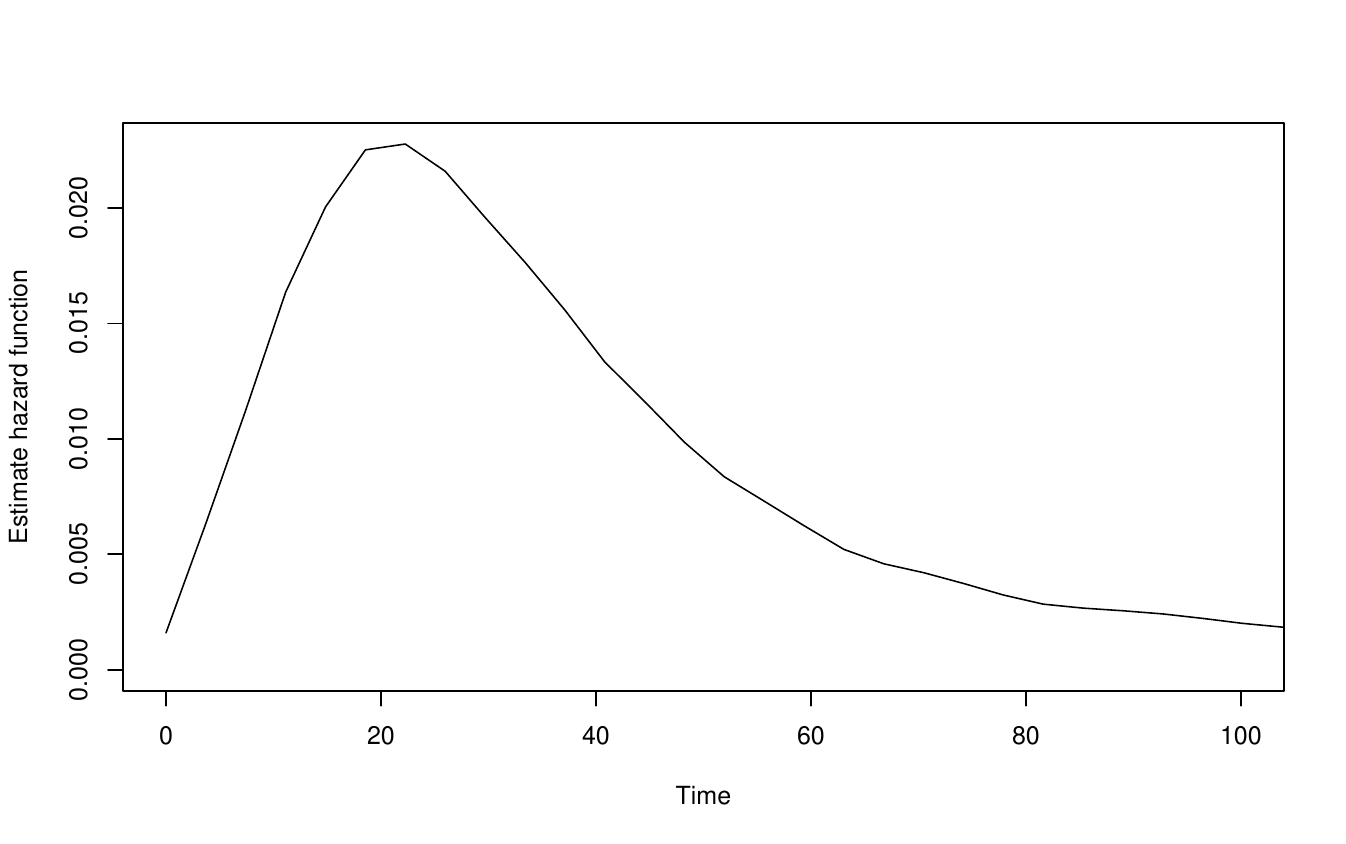} \\
		\caption{Hazard function nonparametric estimate of pregnant and puerperal women hospitalized with SARS due to COVID-19.}
		\label{fig_HF}
	\end{center}
\end{figure}

Figure \ref{fig_HF} presents the hazard function estimated nonparametrically using kernel-based methods \cite{muller1994hazard}. We can observe the estimated hazard function is unimodal, which motivates the consideration of the Dagum distribution. 

 The DD-PVF frailty regression model considered covariate linked to the parameters $\alpha$, $\beta$, and $p_0$. Thus, 
\begin{equation*}
\begin{array}{r}
\label{alpha_application}
\alpha=\exp \left(\zeta_{0}+\zeta_{1} \times \text { age}_\text{$\geq$30}+\zeta_{2} \times \text { group }_\text{postpartum}+\zeta_{3} \times \text { obesity }_\text{yes}+ \right. \\
\left.\zeta_{4} \times \text { loss of taste }_\text{yes}+\zeta_{5} \times \text { saturation }_\text{yes}+\zeta_{6} \times \text { vaccine }_\text{yes}\right),
\end{array}
\end{equation*}

\begin{equation*}
\begin{array}{r}
\label{beta_application}
\beta=\exp \left(-\eta_{0}-\eta_{1} \times \text { age}_\text{$\geq$30}-\eta_{2} \times \text { group }_\text{postpartum}-\eta_{3} \times \text { obesity }_\text{yes}- \right. \\
\left.\eta_{4} \times \text { loss of taste }_\text{yes}-\eta_{5} \times \text { saturation }_\text{yes}-\eta_{6} \times \text { vaccine }_\text{yes}\right),
\end{array}
\end{equation*}

and
\begin{equation*}
\begin{array}{r} 
\label{p0_aplic}
\log\left(\frac{p_{0}}{1-p_{0}}\right)=\nu_{0}+\nu_{1} \times \text { age}_\text{$\geq$30}+\nu_{2} \times \text { group }_\text{postpartum}+\nu_{3} \times \text { obesity }_\text{yes}+  \\
 \nu_{4} \times \text { loss of taste }_\text{yes}+\nu_{5} \times \text { saturation }_\text{yes}+\nu_{6} \times \text { vaccine }_\text{yes}.
\end{array}
\end{equation*}

The results of the fitted DD-PVF frailty model and defective Dagum distribution without frailty  (DD model) are given in Table \ref{table_estimates}. Notice that the estimate of $\gamma$ is close to zero indicating that a DD-gamma frailty model can be considered. In this sense, we also fitted the main special cases, DD-inverse Gaussian ($\gamma = 0.5$) and gamma ($\gamma \rightarrow 0$) frailty models. According to the AIC value, the DD-gamma frailty model seems to be the best choice among the four models. 

The final DD-gamma model is given by:
\begin{equation*}
\begin{array}{r}
\label{alpha_application1}
\alpha=\exp \left(\zeta_{0}+\zeta_{1} \times \text { group }_\text{postpartum}+\zeta_{2} \times \text { desaturation }_\text{yes}\right),
\end{array}
\end{equation*}

\begin{equation*}
\begin{array}{r}
\label{beta_application1}
\beta=\exp \left(-\eta_{0}-\eta_{1} \times \text { age}_\text{$\geq$30}-\eta_{2} \times \text { group }_\text{postpartum}-\eta_{3} \times \text { obesity }_\text{yes} \right. \\
\left. - \eta_{4} \times \text { saturation }_\text{yes}-\eta_{5} \times \text { vaccine }_\text{yes}\right),
\end{array}
\end{equation*}

and
\begin{equation}
\begin{array}{r} 
\label{p0_aplic}
\log\left(\frac{p_{0}}{1-p_{0}}\right)=\nu_{0}+\nu_{1} \times \text { group }_\text{postpartum}+ \nu_{2} \times \text { saturation }_\text{yes}+\nu_{3} \times \text { vaccine }_\text{yes}.
\end{array}
\end{equation}

The estimates of the fitted model \eqref{p0_aplic} are presented in Table \ref{table_estimates_final}. Group and desaturation are the only covariates for $\alpha$. It is important to observe that for all combinations of group and desaturation status, the $\alpha$ values are greater than 1 - for pregnant without desaturation, $\alpha = 2.231$; for pregnant with desaturation, $\alpha = 1.926$; for postpartum without desaturation, $\alpha = 1.949$; for postpartum with desaturation, $\alpha =  1.682$ - it indicates that the hazard function has unimodal form for all them. 

\begin{figure}[h!]
	\centering
	\begin{center}
		\includegraphics[scale = 0.7]{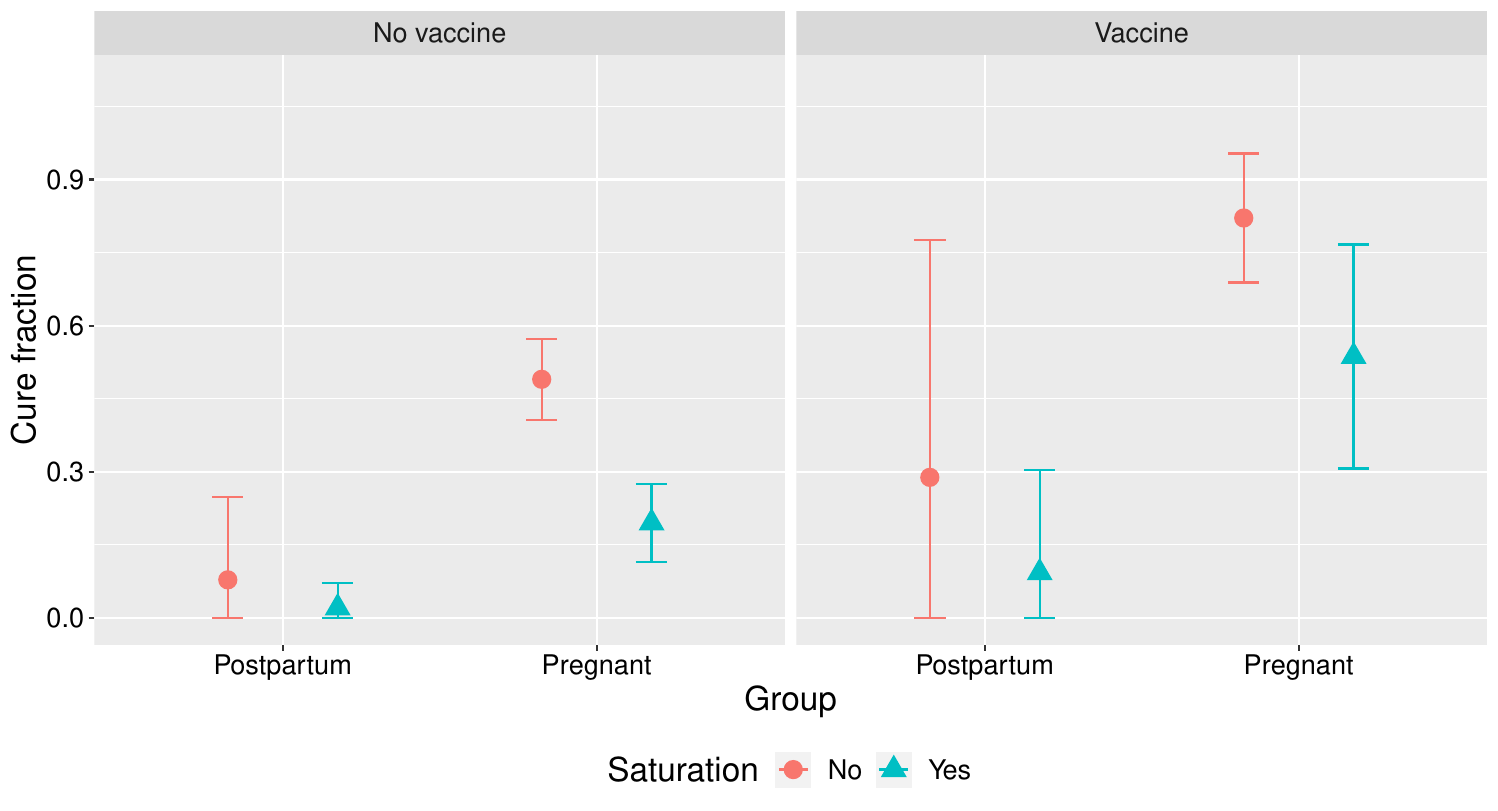} \\
		\caption{Cure fraction estimate for pregnant and puerperal women hospitalized with SARS due to COVID-19.}
		\label{fig_cure-fraction}
	\end{center}
\end{figure}

Only loss of taste does not remain as a predictor for the $\beta$ parameter and, only group, desaturation, and vaccine are important for the cure fraction. Figure \ref{fig_cure-fraction} presents the cure fraction estimates for each combination of group, desaturation, and vaccine status. A pregnant vaccinated woman without desaturation symptoms has the highest cure fraction while a not-vaccinated postpartum woman with desaturation has the lowest cure fraction. Table \ref{table_estimates_theta} presents the $\theta$ estimates for each combination of group, desaturation, and vaccine status. One can note that 1 is inside the 95\% confidence intervals for $\theta$ for the following combinations:  pregnant, desaturation, and no vaccine, and for all combinations with the postpartum group, regardless of desaturation and vaccine status. 

\begin{figure}[h!]
	\centering
	\begin{center}
		\includegraphics[scale = 0.75]{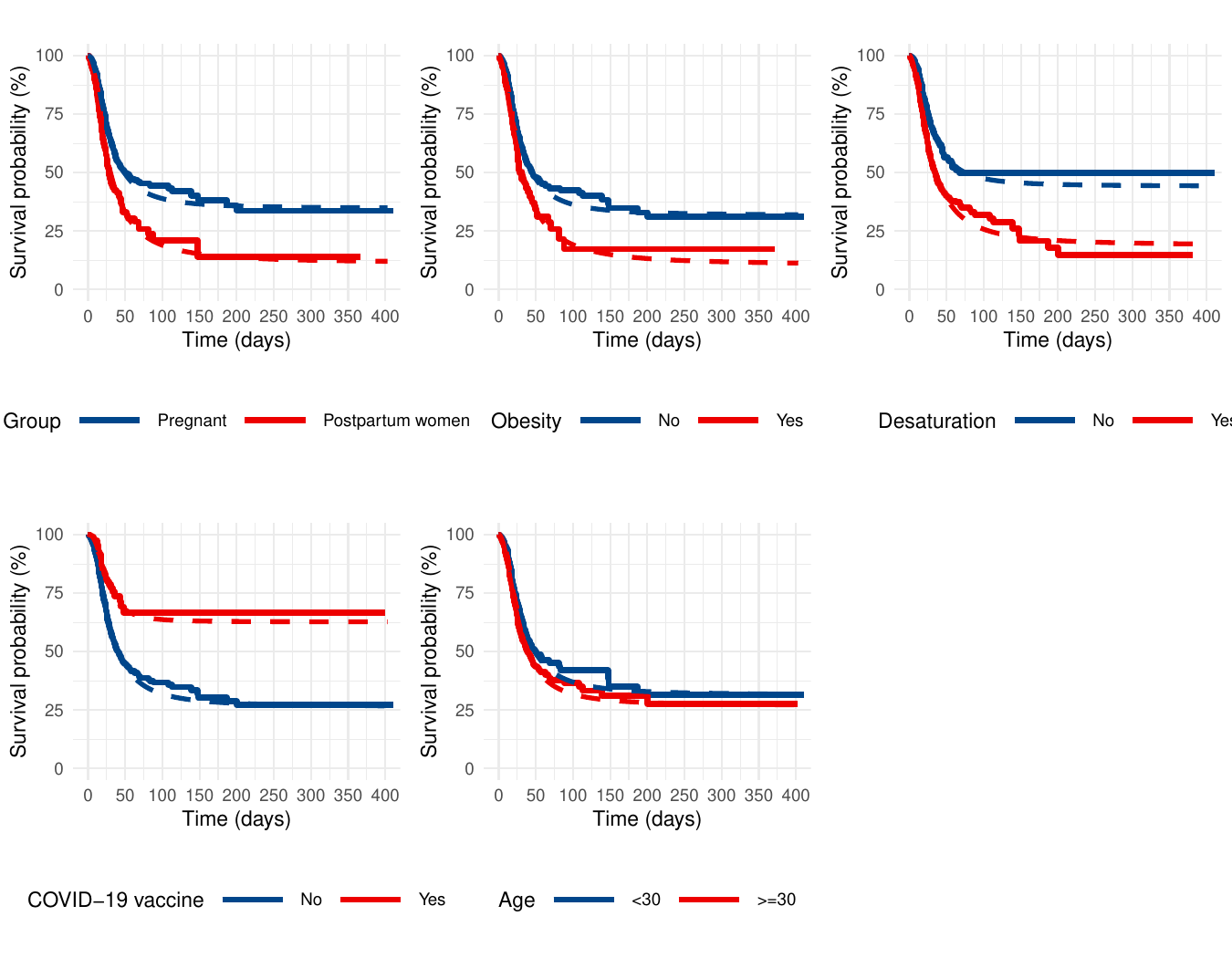} \\
		\caption{Kaplan Meier estimates (solid curves) and DD-PVF model (dashed curves) estimates for survival curves of pregnant and puerperal women hospitalized with SARS due to COVID-19.}
		\label{KMs_combined_fitted-model-maternal}
	\end{center}
\end{figure}

Figure \ref{KMs_combined_fitted-model-maternal} shows the estimated survival functions of the DD-gamma model for each prognostic factor. The survival function estimates by the DD-gamma model are close to the K-M curves.

% \usepackage{multirow}
\begin{table}[h!]
\caption{Parameters estimates for application.}
 \label{table_estimates}
\begin{tabular}{c|cc|cc|cc|cc}
\hline
 & \multicolumn{2}{c|}{DD model}                                      & \multicolumn{2}{c|}{DD - PVF model} & \multicolumn{2}{c|}{DD - Gamma model} & \multicolumn{2}{c}{DD - Inv. Gaussian model} \\
 \hline
\multicolumn{1}{c|}{Parameter}                           & MLE                        & CI 95\%                              & MLE        & CI 95\%               & MLE          & CI 95\%               & MLE              & CI 95\%                   \\
\hline
\multicolumn{1}{c|}{$\sigma^2$}                     & -                          & -                                    & 0.489      & 0.184 - 0.793         & 0.494        & 0.190 -- 0.798        & 0.370            & 0.057 -- 2.407            \\
\multicolumn{1}{c|}{$\gamma$}                      & -                          & -                                    & 0.006      & 0 - 0.136             & -            & -                     & -                & -                         \\
$\zeta_0$                                          & \multicolumn{1}{|c}{0.802}  & \multicolumn{1}{c|}{0.701 -- 0.903}   & 0.810      & 0.714 -- 0.906        & 0.811        & 0.714 -- 0.907        & 0.813            & 0.714 -- 0.912            \\
$\zeta_1(age)$                                         & \multicolumn{1}{|c}{0.008}  & \multicolumn{1}{c|}{-0.093 -- 0.108}  & 0.003      & -0.094 -- 0.099       & 0.004        & -0.093 -- 0.100       & 0.002            & -0.097 -- 0.101           \\
$\zeta_2(group)$                                          & \multicolumn{1}{|c}{-0.159} & \multicolumn{1}{c|}{-0.264 -- -0.053} & -0.113     & -0.214 -- -0.013      & -0.114       & -0.214 -- -0.013      & -0.129           & -0.248 -- -0.009          \\
$\zeta_3(obesity)$                                        & \multicolumn{1}{|c}{-0.047} & \multicolumn{1}{c|}{-0.182 -- 0.088}  & -0.030     & -0.158 -- 0.097       & -0.028       & -0.156 -- 0.099       & -0.042           & -0.172 -- 0.088           \\
$\zeta_4(loss-taste)$                                          & \multicolumn{1}{|c}{0.029}  & \multicolumn{1}{c|}{-0.136 -- 0.195}  & 0.006      & -0.143 -- 0.155       & 0.009        & -0.140 -- 0.158       & 0.009            & -0.153 -- 0.170           \\
$\zeta_5(desaturation)$                                          & \multicolumn{1}{|c}{-0.161} & \multicolumn{1}{c|}{-0.270 -- -0.052} & -0.160     & -0.262 -- -0.057      & -0.161       & -0.264 -- -0.058      & -0.167           & -0.274 -- -0.059          \\
$\zeta_6(vaccine)$                                          & \multicolumn{1}{|c}{0.240}  & \multicolumn{1}{c|}{-0.038 -- 0.517}  & 0.177      & -0.086 -- 0.440       & 0.167        & -0.096 -- 0.431       & 0.172            & -0.161 -- 0.505           \\
$\eta_0$                                            & 8.175                      & 7.621 -- 8.728                       & 8.235      & 7.682 -- 8.789        & 8.240        & 7.686 -- 8.795        & 8.238            & 7.676 -- 8.801            \\
$\eta_1(age)$                                            & -0.232                     & -0.758 -- 0.294                      & -0.276     & -0.799 -- 0.247       & -0.272       & -0.795 -- 0.251       & -0.267           & -0.795 -- 0.261           \\
$\eta_2(group)$                                            & -1.320                     & -1.844 -- -0.797                     & -1.179     & -1.694 -- -0.664      & -1.182       & -1.697 -- -0.667      & -1.229           & -1.790 -- -0.668          \\
$\eta_3(obesity)$                                           & -0.529                     & -1.230 -- 0.171                      & -0.495     & -1.186 -- 0.195       & -0.487       & -1.178 -- 0.205       & -0.532           & -1.224 -- 0.160           \\
$\eta_4(loss-taste)$                                           & 0.410                      & -0.480 -- 1.300                      & 0.324      & -0.511 -- 1.160       & 0.343        & -0.494 -- 1.180       & 0.333            & -0.537 -- 1.202           \\
$\eta_5(desaturation)$                                        & -1.403                     & -1.989 -- -0.816                     & -1.424     & -1.995 -- -0.853      & -1.432       & -2.003 -- -0.860      & -1.447           & -2.037 -- -0.857          \\
$\eta_6(vaccine)$                                            & 2.244                      & 0.442 -- 4.045                       & 1.986      & 0.297 -- 3.676        & 1.922        & 0.246 -- 3.599        & 1.924            & -0.015 -- 3.863           \\
$\nu_0$                                             & 0.214                      & -0.316 -- 0.745                      & 0.445      & -0.020 -- 0.910       & 0.440        & -0.026 -- 0.906       & 0.397            & -0.245 -- 1.039           \\
$\nu_1(age)$                                             & -0.296                     & -0.832 -- 0.240                      & -0.490     & -1.051 -- 0.071       & -0.484       & -1.045 -- 0.078       & -0.434           & -1.068 -- 0.200           \\
$\nu_2(group)$                                              & -1.828                     & -3.073 -- -0.584                     & -6.025     & -43.674 -- 31.623     & -4.558       & -9.257 -- 0.141       & -2.375           & -5.351 -- 0.600           \\
$\nu_3(obesity)$                                             & -1.005                     & -2.149 -- 0.139                      & -1.850     & -3.599 -- -0.100      & -1.819       & -3.532 -- -0.106      & -1.618           & -4.583 -- 1.346           \\
$\nu_4(loss-taste)$                                             & -0.038                     & -1.014 -- 0.938                      & -0.557     & -1.818 -- 0.704       & -0.550       & -1.804 -- 0.705       & -0.364           & -1.993 -- 1.266           \\
$\nu_5(desaturation)$                                             & -1.231                     & -1.806 -- -0.655                     & -1.761     & -2.547 -- -0.974      & -1.778       & -2.579 -- -0.976      & -1.562           & -2.726 -- -0.399          \\
$\nu_6(vaccine)$                                             & 1.803                      & 1.055 -- 2.552                       & 2.108      & 1.191 -- 3.024        & 2.113        & 1.191 -- 3.034        & 1.869            & 0.961 -- 2.777            \\
\hline
AIC                                            & \multicolumn{2}{c|}{12447.84}                                      & \multicolumn{2}{c|}{12448.53}       & \multicolumn{2}{c|}{12446.58}         & \multicolumn{2}{c}{12449.03}                 \\
maxL                                           & \multicolumn{2}{c|}{-6202.921}                                     & \multicolumn{2}{c|}{-6201.27}       & \multicolumn{2}{c|}{-6201.289}        & \multicolumn{2}{c}{-6202.51}    \\
\hline
\end{tabular}
\end{table}

\begin{table}[h!]
\caption{Parameters estimates for application. Final model}
 \label{table_estimates_final}
 \centering
\begin{tabular}{c|ccc}
\hline
 Parameter &  MLE & \multicolumn{2}{c}{CI 95\%} \\
 \hline 
 \multicolumn{4}{c}{$\alpha$}                                             \\
 \hline
$\zeta_0$                         & 0.803                   & 0.718        & 0.887        \\
$\zeta_1(group)$                        & -0.135                  & -0.246       & -0.024       \\
$\zeta_2(desaturation)$                        & -0.147                  & -0.249       & -0.045       \\
 \hline 
 \multicolumn{4}{c}{$\beta$} \\
 \hline
$\eta_0$                         & 8.260                   & 7.764        & 8.755        \\
$\eta_1(age)$                          & -0.321                  & -0.484       & -0.157       \\
$\eta_2(group)$                           & -1.265                  & -1.803       & -0.727       \\
$\eta_3(obesity)$                           & -0.343                  & -0.573       & -0.113       \\
$\eta_4(desaturation)$                        & -1.362                  & -1.919       & -0.804       \\
$\eta_5(vaccine)$                          & 0.921                   & 0.479        & 1.363        \\
 \hline 
 \multicolumn{4}{c}{$p_0$} \\
 \hline
$\nu_0$                            & -0.040                  & -0.374       & 0.293        \\
$\nu_1(group)$                            & -2.425                  & -4.813       & -0.037       \\
$\nu_2(desaturation)$                            & -1.375                  & -1.971       & -0.779       \\
$\nu_3(vaccine)$                             & 1.563                   & 0.663        & 2.463        \\
\hline
$\sigma^2$                         & 0.294                   & 0.082        & 1.056    \\
\hline
\end{tabular}
\end{table}

\begin{table}[h!]
\centering
\caption{$\theta$ estimates}
 \label{table_estimates_theta}
\begin{tabular}{ccc|ccc}
\hline
Group      & Desaturation & Vaccine & MLE   & \multicolumn{2}{c}{CI 95\%} \\
\hline
Pregnant   & No           & No      & 0.548 & 0.441        & 0.655        \\
Postpartum & No           & No      & 0.977 & 0.857        & 1.098        \\
Pregnant   & Yes          & No      & 0.877 & 0.748        & 1.007        \\
Postpartum & Yes          & No      & 0.999 & 0.991        & 1.007        \\
Pregnant   & No           & Vaccine & 0.184 & 0.045        & 0.323        \\
Postpartum & No           & Vaccine & 0.777 & 0.217        & 1.337        \\
Pregnant   & Yes          & Vaccine & 0.495 & 0.229        & 0.760        \\
Postpartum & Yes          & Vaccine & 0.968 & 0.803        & 1.133       \\
\hline
\end{tabular}
\end{table}

\subsection{Malignant neoplasms of skin (not melanoma)} \label{application_fosp}
 
Skin cancers, with their diverse manifestations, have garnered significant attention in the field of oncology. While melanoma and non-melanoma skin cancers have been extensively studied, the category ``Other malignant neoplasms of the skin'' remains relatively less explored. This group encapsulates a myriad of rare and enigmatic malignancies, each characterized by distinct clinical presentations, histopathology attributes, and therapeutic considerations. This type of skin cancer is classified as C44 in the ICD-10 (International Classification of Diseases, 10th Revision), which is a system for categorizing and coding diseases, health conditions, and related factors.

The epidemiology of C44 is marked by its rarity, with varied incidences across different subtypes. In contrast, melanoma, particularly cutaneous melanoma, commands a higher incidence and is more widely recognized. Diverse etiological factors contribute to C44. Ultraviolet radiation exposure level is a key player in the C44 and displays associations with immunosuppression and viral infections. However, this type of information is not generally available and is lacking in the study of survival for C44 patients.

In this application, the goal is to estimate the survival and the cure fraction of patients with other malignant neoplasms of the skin, specifically the skin of the lower limb (including the hip), which englobes the ICD-10 category C44.7.  This information would allow physicians to prevent the overall progressive burden of this disease, with measures of control and preventive interventions in this context.

For that, we consider a malignant neoplasm of the skin dataset from a retrospective survey of $14193$ records of patients diagnosed with malignant neoplasms of the skin of the lower limb in the state of S\~ao Paulo, Brazil, between 2000 and 2023, with follow-up conducted until June of 2023 and with at least two months of follow-up.  Death due to C44.7 cancer was defined as the event of interest. Those patients who did not die due to cancer during the follow-up period were characterized as right-censored observations. In this analysis, we want to study the impact of the metastasis stage at the diagnosis in survival of skin cancer patients, in order to understand how the disease affects this population.  

This first-time analyzed dataset is provided by the Fundaç\~ao Oncocentro de S\~ao Paulo (FOSP), which is responsible for coordinating and monitoring the implementation of the Hospital Cancer Registry in the State of S\~ao Paulo (Brazil), in addition to systematizing and evaluating cancer care data available for the state. The FOSP is a public institution connected to the State Health Secretariat that monitors the evolution of the Oncology Care Network, assists the State Department of Health in the creation and application of prevention and health promotion programs, and monitors the evolution of cancer mortality in the state.  

\begin{figure}[h!]
	\centering
	\begin{center}
		\includegraphics[scale = 0.4]{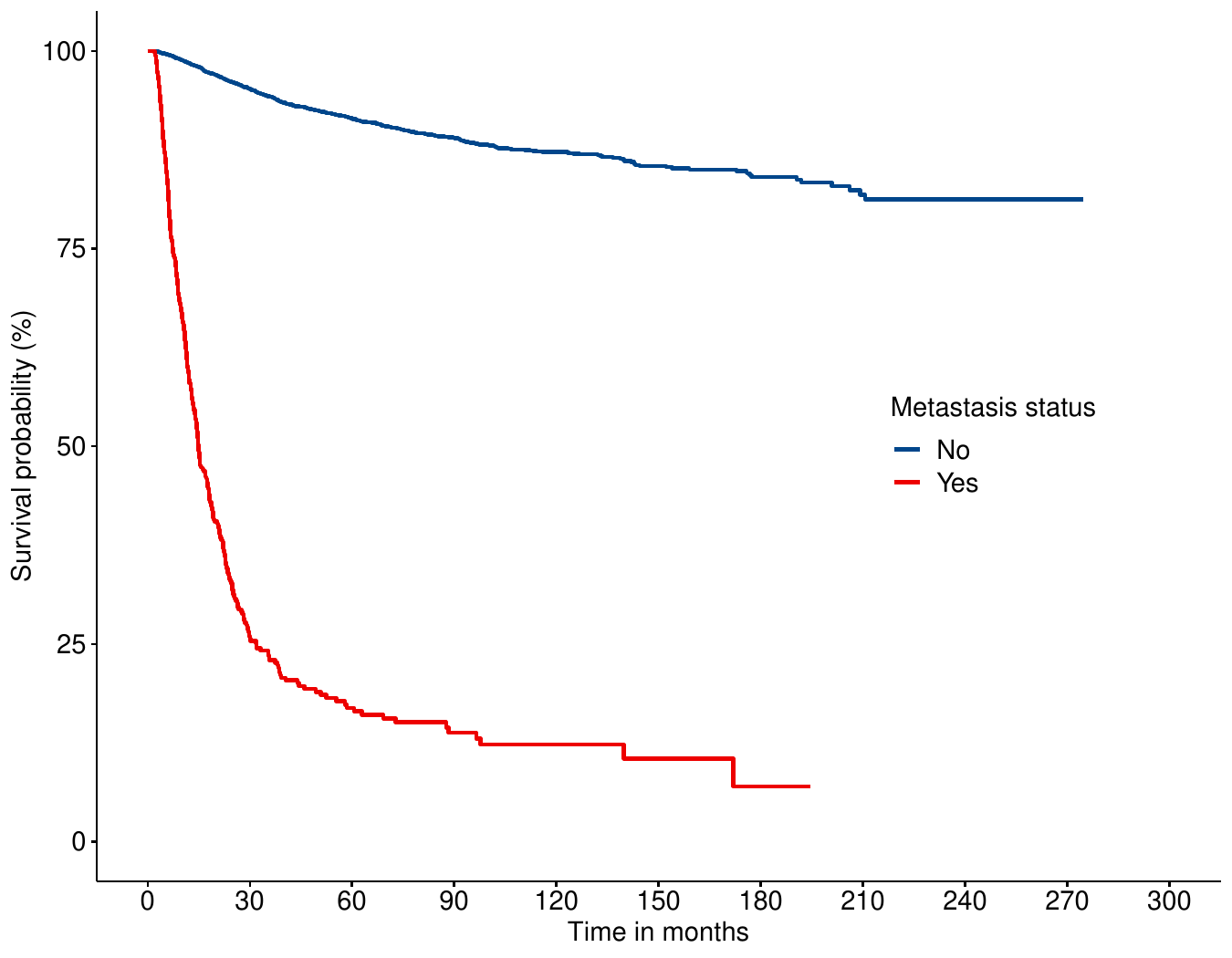} \\
		\caption{Kaplan Meier estimates for survival curves of malignant skin cancer (not melanoma} patients.
		\label{fig_KM_skin}
	\end{center}
\end{figure}

Of the $14193$ patients, $91\%$ did not die during the follow-up period, that is, $12950$ patients have a right-censored event time. A total of $13694$ (96\%) patients are not in the metastasis stage and $499$ (4\%) are in the metastasis stage at the diagnosis moment. Figure \ref{fig_KM_skin} presents the Kaplan-Meier estimates grouped by metastasis status. There is strong evidence of a cured fraction for the no-metastasis group, while the metastasis group has no long-term survivors. It is worth mentioning that only $6.3\%$ ($876/13694)$ of the no-metastasis group present the event of interest during the follow-up and this percentage grows up to $73.5\%$ ($367/499)$ for the metastasis group. 

\begin{figure}[h!]
	\centering
	\begin{center} 
  		\includegraphics[scale = 0.6]{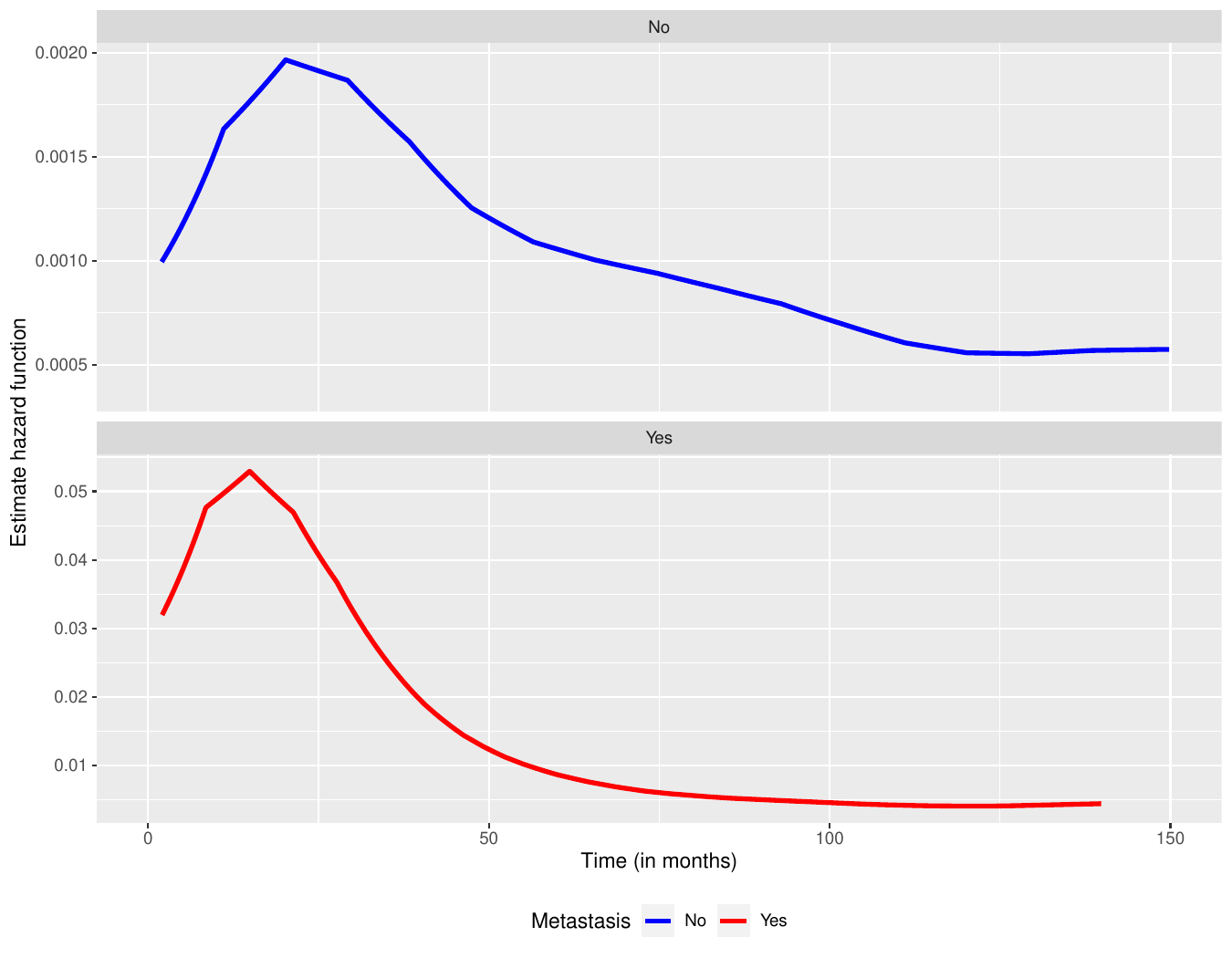} \\
		\caption{Hazard function non-parametric estimate of malignant neoplasms of skin.}
		\label{fig_HF_cancer_skin}
	\end{center}
\end{figure}

Figure \ref{fig_HF_cancer_skin} presents the hazard function estimated nonparametrically using kernel-based methods \cite{muller1994hazard}. We can observe the estimated hazard function is unimodal, which motivates considering the proposed model. 

The results of the fitted DD-PVF, DD-gamma, DD-inverse Gaussian, and DD models are given in Table \ref{table_estimates_skin}. Based on AIC values, DD-PVF model seems to be the best choice among the four models, and the estimated value for the $\gamma$ parameter is $0.727$ (CI $95\% 0.631 -- 0.824$). The maximum profile likelihood estimates were also obtained for this dataset and the results are similar to the maximum likelihood approach.

The metastasis stage is significant when considered for $\alpha$, $\beta$, and $p_0$ parameters. The estimates of $p_0$ are presented in Table \ref{table_estimates_p0_skin} and one can observe that for the no-metastasis group, the cure fraction is estimated in $81.7\%$, while it is close to zero (4.9\%) for the group in the metastasis stage. For this last group, 1 is inside the 95\% confidence interval for $\theta$, which indicates there is no cure fraction for this subpopulation (Table \ref{table_estimates_theta_skin}).

\begin{figure}[h!]
	\centering
	\begin{center}
		\includegraphics[scale = 0.6]{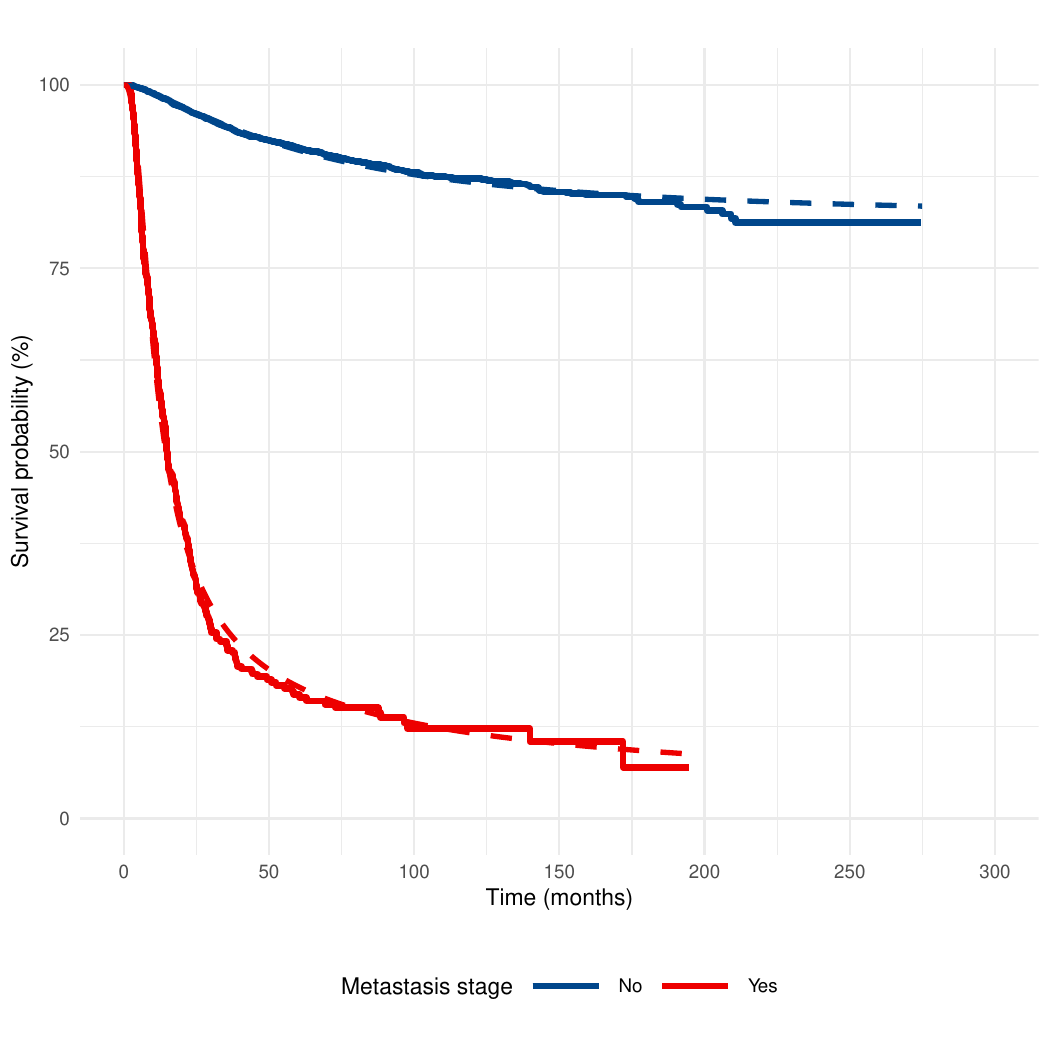} \\
		\caption{Kaplan Meier estimates (solid curves) and DD-PVF model (dashed curves) estimates for survival curves of malignant skin cancer (not melanoma) patients.}
		\label{KMs_combined_fitted-model-skin}
	\end{center}
\end{figure}

Figure \ref{KMs_combined_fitted-model-skin} shows the estimated survival functions of the DD-PVF model for each metastasis stage. The survival function estimates by the DD-PVF model are close to the K-M curves. 

\begin{table}[h!]
\caption{Parameters estimates for the malignant neoplasms of skin application.}
 \label{table_estimates_skin}
\begin{tabular}{c|cc|cc|cc|cc}
\hline
 & \multicolumn{2}{c|}{DD model}                                      & \multicolumn{2}{c|}{DD - PVF model} & \multicolumn{2}{c|}{DD - Gamma model} & \multicolumn{2}{c}{DD - Inv. Gaussian model} \\
 \hline
\multicolumn{1}{c|}{Parameter}                           & MLE                        & CI 95\%                              & MLE        & CI 95\%               & MLE          & CI 95\%               & MLE              & CI 95\%                   \\
\hline
\multicolumn{1}{c|}{$\sigma^2$}                     & -                          & -                                    &  11.920 & 2.869 -- 49.517
         & 0.735        & 0.508 --	1.063      & 1.766            & 0.975 --	3.196
            \\
\multicolumn{1}{c|}{$\gamma$}                      & -                          & -                                    & 0.727
      & 0.631 -- 0.824  & -            & -                     & -                & -                         \\
\hline
 \multicolumn{9}{c}{$\alpha$}     \\
 \hline
$\zeta_0$                                          & \multicolumn{1}{|c}{0.401}  & \multicolumn{1}{c|}{0.326 -- 0.476}   & 0.449
      & 0.357 --	0.540
        & 0.401
       & 0.326	-- 0.476
       & 0.404    & 0.329 --	0.479
           \\
$\zeta_1(metastasis)$                                         & \multicolumn{1}{|c}{0.254}  & \multicolumn{1}{c|}{0.132 --	0.376}  & 0.642
      & 0.462 --	0.821
      & 0.412        & 0.263 --	0.561
      & 0.527           & 0.341 --	0.712
          \\
\hline
 \multicolumn{9}{c}{$\beta$}     \\
 \hline
$\eta_0$                                            & 7.854                      & 7.489	-- 8.220 & 7.979
      & 7.586	-- 8.372
        & 7.854       & 7.488 --	8.219
       & 7.861            & 7.496 --	8.227
           \\
$\eta_1(metastasis)$                                            & -2.766
                     & -3.366 --	-2.166
                    & -1.505
    & -2.464 --	-0.545
       & -2.201
       & -2.924 -- -1.478
     & -1.804          & -2.677	--  -0.931
        \\
\hline
 \multicolumn{9}{c}{$p_0$}     \\
 \hline
$\nu_0$                                             & 1.513                      & 1.366 --	1.661
                      & 1.499
      & 1.345	-- 1.653
      & 1.513
        & 1.365 --	1.660
      & 1.509           & 1.360 -- 1.658
         \\
$\nu_1(metastasis)$                                             & -3.598                     & -4.034	-- -3.161                     & -4.458
    & -8.457 --	-0.460
      & -4.884     & -28.841 --	19.072
     & -5.149         & -34.587	-- 24.289
         \\
\hline
AIC                                            & \multicolumn{2}{c|}{16350.63}                                      & \multicolumn{2}{c|}{16338.46}       & \multicolumn{2}{c|}{16344.25}         & \multicolumn{2}{c}{16341.38}                 \\
maxL                                           & \multicolumn{2}{c|}{-8169.315}                                     & \multicolumn{2}{c|}{-8161.23}       & \multicolumn{2}{c|}{-8165.124}        & \multicolumn{2}{c}{-8163.69}    \\
\hline
\end{tabular}
\end{table}

\begin{table}[h!]
\centering
\caption{Cure fraction ($p_0$) estimates for the malignant neoplasms of skin application.}
 \label{table_estimates_p0_skin}
\begin{tabular}{c|ccc}
\hline
Metastasis     & MLE   & \multicolumn{2}{c}{CI 95\%} \\
\hline
No   &  0.817  & 0.794 &	0.840      \\
Yes &  0.049  & 0        & 0.237    \\
\hline
\end{tabular}
\end{table}

\begin{table}[h!]
\centering
\caption{$\theta$ estimates for the malignant neoplasms of skin application.}
 \label{table_estimates_theta_skin}
\begin{tabular}{c|ccc}
\hline
Metastasis     & MLE   & \multicolumn{2}{c}{CI 95\%} \\
\hline
No   &  0.286  & 0.204 &	0.368      \\
Yes &  0.999  & 0.998     & 1.001    \\
\hline
\end{tabular}
\end{table}

\clearpage 

\section{Conclusions}\label{final_considerations}

In this paper, we introduced the Defective Dagum-PVF (DD-PVF) regression model, an extension of the conventional Dagum distribution. It incorporates a power variance function (PVF) frailty term within the hazard function in a multiplicative way. This added flexibility is particularly valuable when dealing with cure fractions and non-monotonic hazard functions.

The DD-PVF model is highlighted for allowing a direct interpretation of how covariates influence the cure fraction. This interpretability sets it apart from other defective models. Furthermore, it allows the exploration of scenarios in which one subgroup has long-term survivors, while another does not.

Simulation studies validated the efficacy of maximum likelihood estimators for the DD-PVF model. Comparing the DD-PVF model to its counterpart without the PVF frailty term revealed a significant advantage, especially as the frailty term's variance increases. The model demonstrates a high sensitivity to discriminate between scenarios where one subgroup features long-term survivors and the other does not.

In conclusion, our work underscores the compelling reasons and remarkable benefits of the DD-PVF model. Through its rigorous application to two authentic datasets, we have illuminated the model's practical relevance. One dataset scrutinized severe COVID-19 cases among pregnant and postpartum women in Brazil, while the other probed patients with malignant skin neoplasms. These insightful applications underscore the DD-PVF model's valuable contributions to the field of survival analysis and motivate its continued exploration in diverse clinical and epidemiological contexts.

%\section*{Acknowledgments}
\section*{Appendix A. Proof of Theorem 1}
\renewcommand{\theequation}{A\arabic{equation}}
\setcounter{equation}{0}
As $S_0(t)$ is a survival function of a defective distribution, then $\lim_{t\rightarrow\infty}S_0(t)=p_0 \in (0,1)$. So, 
\begin{eqnarray*}
\lim_{t\rightarrow\infty} S(t\mid \gamma,\sigma^2)=\exp\left(\frac{1-\gamma}{\gamma \sigma^2}\left\{1-\left[1-\frac{\sigma^2 \log(p_0)}{1-\gamma}\right]^{\gamma}\right\}\right).
\end{eqnarray*}
As $p_0 \in (0,1)$, $\sigma^2>0$ and $0<\gamma<1$, then
\begin{eqnarray}
\displaystyle \frac{\sigma^2}{1-\gamma}\log(p_0) <0 \Leftrightarrow -\frac{\sigma^2}{1-\gamma}\log(p_0) > 0 \Leftrightarrow 1-\frac{\sigma^2}{1-\gamma}\log(p_0) > 1 \Leftrightarrow   \nonumber \\
\left[1-\frac{\sigma^2}{1-\gamma}\log(p_0)\right]^{\gamma}  > 1 \Leftrightarrow -\left[1-\frac{\sigma^2}{1-\gamma}\log(p_0)\right]^{\gamma}  < -1  \Leftrightarrow\nonumber \\ 
1-\left[1-\frac{\sigma^2}{1-\gamma}\log(p_0)\right]^{\gamma}  < 0 \Leftrightarrow \frac{1-\gamma}{\gamma\sigma^2}\left\{1-\left[1-\frac{\sigma^2}{1-\gamma}\log(p_0)\right]^{\gamma}\right\}  < 0  \Leftrightarrow \nonumber \\
\exp\left(\frac{1-\gamma}{\gamma\sigma^2}\left\{1-\left[1-\frac{\sigma^2}{1-\gamma}\log(p_0)\right]^{\gamma}\right\}\right) < 1. \nonumber
\end{eqnarray}
Besides, as $\frac{1-\gamma}{\gamma\sigma^2}\left\{1-\left[1-\frac{\sigma^2}{1-\gamma}\log(p_0)\right]^{\gamma}\right\} < 0$ and $\lim_{x\rightarrow-\infty}\exp(x)=0$, thus
\begin{eqnarray}
0 < \exp\left(\frac{1-\gamma}{\gamma\sigma^2}\left\{1-\left[1-\frac{\sigma^2}{1-\gamma}\log(p_0)\right]^{\gamma}\right\}\right) < 1, \nonumber
\end{eqnarray}
Therefore, if $\frac{\sigma^2}{1-\gamma} \log[S_0(t)]<1$, 
\begin{eqnarray*}
\lim_{t\rightarrow\infty} S(t\mid \gamma,\sigma^2)=\exp\left(\frac{1-\gamma}{\gamma \sigma^2}\left\{1-\left[1-\frac{\sigma^2 \log(p_0)}{1-\gamma}\right]^{\gamma}\right\}\right) \in (0,1).
\end{eqnarray*}

%\subsection*{Author contributions}

%\subsection*{Financial disclosure}

%None reported.

\subsection*{Conflict of interest}

The authors declare no potential conflict of interests.

%\section*{Supporting information}

%The following supporting information is available as part of the online article:

\clearpage
 
\bibliography{references}%

\clearpage

%\section*{Author Biography}

%\begin{biography}{\includegraphics[width=66pt,height=86pt,draft]{empty}}{\textbf{Author Name.} This is sample author biography text this is sample author biography text this is sample author biography text this is sample author biography text this is sample author biography text this is sample author biography text this is sample author biography text this is sample author biography text this is sample author biography text this is sample author biography text this is sample author biography text this is sample author biography text this is sample author biography text this is sample author biography text this is sample author biography text this is sample author biography text this is sample author biography text this is sample author biography text this is sample author biography text this is sample author biography text this is sample author biography text.}
%\end{biography}

\end{document}